\documentclass[12pt]{article}
\pdfoutput=1

\usepackage{cancel,slashed}
\usepackage{bbm}
\usepackage{amssymb}
\usepackage{amsfonts}
\usepackage{amsmath}
\usepackage{graphicx}
\usepackage{cite}

\usepackage{latexsym}
\usepackage{color}
\usepackage{xypic}
\usepackage{cancel,slashed}
\usepackage[hyperfootnotes=false,linktocpage]{hyperref}
\usepackage{color}
\usepackage{transparent}
\usepackage[hang,flushmargin]{footmisc}

\usepackage{blkarray}
\usepackage{multirow}

\makeatletter
\def\@xfootnote[#1]{%
  \protected@xdef\@thefnmark{#1}%
  \@footnotemark\@footnotetext}
\makeatother

\newcommand{\bmat}{\left(\begin{array}}
\newcommand{\emat}{\end{array}\right)}

\def\p{\partial}
\def\a{\alpha}

\def\b{\beta}
\def\g{\gamma}
\def\d{\delta}
\def\th{\theta}
\def\om{\omega}
\def\Om{\Omega}

\def\-{\hphantom{-}}
\def\ov{\overline}
\def\s2{\frac{1}{\sqrt2}}

\def\oh{\frac{1}{2}}
\def\beq{\begin{equation}}
\def\eeq{\end{equation}}
\def\beqa{\begin{eqnarray}}
\def\eeqa{\end{eqnarray}}

\def\im{{\rm Im \,}}
\def\re{{\rm Re \,}}

\def\Z{{\mathbb Z}}

\def\cc{{\mathcal C}}

\def\cg{{\mathcal G}}

\def\cn{{\mathcal N}}

\def\cs{{\mathcal S}}

\def\Dsl{\,\raise.15ex\hbox{/}\mkern-13.5mu D} 

\def\CH {{\cal H}}
\def\CK {{\cal K}}
\def\CM {{\cal M}}
\def\CR {{\cal R}}
\def\CN {{\cal N}}
\def\CF {{\cal F}}
\def\CL {{\cal L}}

\def\CQ {{\cal Q}}
\def\re{\mbox{Re}}
\def\im{\mbox{Im}}

\def\be{\begin{equation}}
\def\ee{\end{equation}}
\def\bea{\begin{eqnarray}}
\def\eea{\end{eqnarray}}
\def\raw{\rightarrow}
\def\Raw{\Rightarrow}

\def\IZ{\mathbb{Z}}
\def\IR{\mathbb{R}}

\def\Id{{\mathbb{I}}}

\def\oh{\frac{1}{2}}
\def\a{{\alpha}}
\def\b{{\beta}}
\def\d{{\delta}}

\def\th{{\theta}}
\def\Lam{{\Lambda}}
\def\lam{{\lambda}}
\def\Om{{\Omega}}
\def\om{{\omega}}
\def\sig{{\sigma}}
\def\Sig{{\Sigma}}
\def\g{{\gamma}}
\def\G{{\Gamma}}

\def\p{{\partial}}

\definecolor{mygr}{rgb}{0,0.6,0}

\def\sm2{{\mbox{\small 2}}}

\definecolor{mygr}{rgb}{0,0.6,0}
\definecolor{mygrey}{rgb}{0,0.1,0.2}
\definecolor{myblue}{rgb}{0,0.5,0.9}
\definecolor{myblue2}{rgb}{0,0.5,0.5}
\definecolor{myorange}{rgb}{1,0.5,0}
\definecolor{mypurple}{rgb}{0.4,0,1}
\definecolor{mygolden}{rgb}{1,0.8,0.2}
\definecolor{mycyan}{rgb}{0,1,1}
\definecolor{mymagenta}{rgb}{1,0,1}

\begin{document}
\pagestyle{plain}

\makeatletter
\@addtoreset{equation}{section}
\makeatother
\renewcommand{\theequation}{\thesection.\arabic{equation}}
\pagestyle{empty}
\rightline{ IFT-UAM/CSIC-16-052}
\vspace{0.5cm}
\begin{center}
\Huge{{Open string multi-branched\\ and K\"ahler potentials}
\\[10mm]}
\large{Federico Carta,\footnote[$\dagger$]{La Caixa-Severo Ochoa Scholar}$^1$ Fernando Marchesano,$^1$ Wieland Staessens$^{1,2}$\\ and Gianluca Zoccarato$^{1,2}$ \\[10mm]}
\small{
${}^1$ Instituto de F\'{\i}sica Te\'orica UAM-CSIC, Cantoblanco, 28049 Madrid, Spain \\[2mm] 
${}^2$ Departamento de F\'{\i}sica Te\'orica, 
Universidad Aut\'onoma de Madrid, 
28049 Madrid, Spain
\\[8mm]} 
\small{\bf Abstract} \\[5mm]
\end{center}
\begin{center}
\begin{minipage}[h]{15.0cm} 

We consider type II string compactifications on Calabi-Yau orientifolds with fluxes and D-branes, and analyse the F-term scalar potential that simultaneously involves closed and open string modes. In type IIA models with D6-branes this potential can be directly computed by integrating out Minkowski three-forms. The result shows a multi-branched structure along the space of lifted open string moduli, in which discrete shifts in special Lagrangian and Wilson line deformations are compensated by changes in the RR flux quanta. The same sort of discrete shift symmetries are present in the superpotential and constrain the K\"ahler potential. As for the latter, inclusion of open string moduli breaks the factorisation between complex structure and K\"ahler moduli spaces. Nevertheless, the 4d K\"ahler metrics display a set of interesting relations that allow to rederive the scalar potential analytically. Similar results hold for type IIB flux compactifications with  D7-brane Wilson lines.

\end{minipage}
\end{center}
\newpage
\setcounter{page}{1}
\pagestyle{plain}
\renewcommand{\thefootnote}{\arabic{footnote}}
\setcounter{footnote}{0}


\tableofcontents


\section{Introduction}
\label{s:intro}

A substantial effort in the literature of 4d string compactifications is devoted to construct models with a rich chiral gauge sector and where most neutral scalars are stabilised at a sufficiently high scale \cite{Ibanez:2012zz,Grana:2005jc,Douglas:2006es,Denef:2007pq,Blumenhagen:2005mu,Blumenhagen:2006ci,Marchesano:2007de,Nilles:2008gq,Maharana:2012tu,Schellekens:2013bpa,Quevedo:2014xia}. Achieving both features simultaneously is a non-trivial task, and probably the most developed models in this sense appear in 
type II orientifold compactifications with either O3/O7 or O6-planes. In this type II scheme the chiral gauge sector arises from space-time filling D-branes, and moduli are fixed by a potential generated by internal background  fluxes and non-perturbative effects \cite{Grana:2005jc,Douglas:2006es,Denef:2007pq}. Remarkably, these two features are usually treated independently, and the presence of the D-branes is ignored when computing the F-term potential that stabilises moduli.\footnote{Except when they source non-perturbative effects or implement a vacuum energy uplifting mechanism.}

This picture is somewhat reversed for other classes of models at weak coupling, like heterotic/type I or type IIB compactifications with O5-planes. There the techniques to achieve full moduli stabilisation are not so well developed, but on the other hand moduli in the gravity and gauge sector are treated on equal footing. In these setups one can see that, in general, the presence of D-branes/vector bundle sectors modifies the moduli stabilisation potential \cite{Jockers:2008pe,Grimm:2008dq,Baumgartl:2008qp,Anderson:2010mh,Anderson:2011ty}. Of course, all these different constructions are related to each other by dualities like mirror symmetry, which has been used to match superpotentials involving open and closed string modes in type IIA compactifications with D6-branes. However, in the cases that have so far been explored the source of open string moduli superpotential in the type IIA side is exclusively due to worldsheet instantons, which are very suppressed in the large volume limit in which RR flux potentials are valid. This somehow sustains the perception that D6-brane moduli are to be treated separately from those moduli entering the flux superpotential, and that they should only be considered at a second stage, after the effect of fluxes on the closed string sector has been taken into account. More generally, open string moduli involving D-brane Wilson lines are presumed to only develop superpotentials from either worldsheet or D-brane instantons, being insensitive to and negligible for the dynamics of the flux generated superpotential. 

Contrary to this expectation, it was pointed out in \cite{Marchesano:2014iea} that D-brane Wilson line moduli can also develop superpotentials at large volume and weak coupling, 
comparable in magnitude to the flux generated superpotential for closed string moduli. In those circumstances, one should analyse the process of moduli stabilisation by treating open and closed string modes simultaneously, considering a scalar potential that includes both at the same time. 

The purpose of this work is to consider type II compactifications where open strings modes (and in particular Wilson lines) enter the potential generated by fluxes on equal footing to the closed string ones. Such a combined description is not only necessary for consistency, but also provides valuable information regarding the $\CN=1$ compactification data. Indeed, global knowledge of the scalar potential and the superpotential provides stringent constraints on the K\"ahler potential for 4d chiral fields. In our case it will allow us to elucidate how closed and open string moduli are interrelated and the presence of shift symmetries in the latter, typically a relevant aspect in building models of inflation \cite{Silverstein:2013wua,Baumann:2014nda,Westphal:2014ana}.

In order to implement our approach we will seek for compactifications with two basic features, which allow to reverse engineer the K\"ahler potential dependence on the open string moduli. First, one should be able to compute the F-term scalar potential without prior knowledge of the K\"ahler or superpotential. Second, one should be able to involve all the open string moduli in the superpotential, which can also be computed independently. We find that type IIA compactifications in Calabi-Yau orientifolds with D6-branes are particularly suitable for this purpose, as they fulfil both requirements when we restrict to tree-level potentials at large compactification volumes. 
Indeed, for type IIA compactifications with background fluxes the tree-level F-term potential for closed strings can be fully computed by integrating out 4d three-forms \cite{Bielleman:2015ina}, and along the lines of \cite{Grimm:2011dx,Kerstan:2011dy} one may generalise this computation to include open string moduli into the potential. Moreover one may always generate an F-term for each D6-brane position and Wilson line moduli, by simply adding an internal worldvolume flux to the D6-brane.

In our analysis we find that a particularly important role is played by a series of discrete symmetries manifest both at the level of the scalar potential and superpotential. Whenever a D-brane field appears in the superpotential, closing a non-trivial loop in open string moduli space is not a symmetry by itself, but it must be accompanied by a compensating shift in the RR flux quanta. This unfolding of the open string moduli space and the corresponding discrete shift symmetries determine to large extent how D-brane moduli enter into the superpotential. More precisely, at the level of approximation in which we are working, we find that the superpotential can be written in the form
\be
W_{\rm open} + W_{\rm closed} (m) \, =\, W_{\rm closed} (\tilde m)
\ee
in which $m$ are the usual RR flux quanta and $\tilde m$ are {\it dressed} fluxes: combinations of flux quanta and open string moduli invariant under the discrete shift symmetries, for which we give a simple geometrical interpretation. A similar statement holds at the level of the flux-generated potential, which displays the multi-branched structure discussed in \cite{Bielleman:2015ina} but now enriched with the open string dependence. 

Analysing periodic directions in open string moduli space proves also illuminating to guess the form of the open-closed K\"ahler potential. This is because one may use such  periodic directions to deduce how open string modes redefine closed strings moduli into new holomorphic variables in the 4d effective theory. Such a redefinition dictates in turn how open strings enter into the K\"ahler potential, from where the educated guess follows. As a direct consequence of this approach we find that open string moduli do not redefine closed string moduli by themselves, but always in combination with other closed string moduli. More precisely, we find that in the presence of open string moduli the complex structure and K\"ahler moduli spaces no longer factorise. This is a well-known effect for type II toroidal orientifolds, which we are now able to generalise to the Calabi-Yau context. Despite the resulting complication for the K\"ahler metrics one can still derive interesting relations among them, thanks to the continuous shift symmetries of the tree-level K\"ahler potential and the fact that it can be expressed in terms of homogeneous functions of real fields. Finally, one can use these relations to show that the F-term scalar potential is indeed reproduced by means of the usual 4d $\CN=1$ supergravity formula.

The rest of the paper is organised as follows. In Section \ref{s:typeIIA} we review type IIA compactifications with D6-branes and fluxes, with special emphasis on the description of the open string moduli space. In Section \ref{s:3forms} we compute the tree-level open-closed scalar potential by direct dimensional reduction, repackaged in the convenient language of Minkowski three-forms. Such a potential displays a set of discrete shift symmetries which in Section \ref{s:supo} are also shown to be present at the level of the superpotential. In Section \ref{s:kahler} we describe how holomorphic variables are redefined in the presence of open string moduli and the implications for the open-closed K\"ahler potential. In Section \ref{s:spot} we use these superpotential and K\"ahler potential to recover the F-term scalar potential of Section \ref{s:3forms} via 4d supergravity. Many of these results also apply to type IIB compactifications with O3/O7-planes and D7-brane Wilson lines, as we discuss in Section \ref{s:typeIIb}. We draw our conclusions and directions for future work in Section \ref{s:conclu}.
 
Several technical details have been relegated to the appendices. Appendix \ref{ap:details} discusses aspects of the open-closed K\"ahler metrics and contains the proof of several identities necessary for the computations of Section \ref{s:spot}. Appendix \ref{ap:dimred} contains a direct derivation of the type IIA flux potential in the democratic formulation of 10d supergravity. Finally, Appendix \ref{ap:texample} illustrates the somewhat abstract definitions used along the main text in the simple case of a ${\bf T}^2\times {\bf T}^4/\Z_2$ orientifold example.

\section{D6-branes in type IIA orientifolds}
\label{s:typeIIA}

Let us consider a type IIA orientifold compactification on $\mathbb R^{1,3}\times \mathcal M_6$ with $\mathcal M_6$ a compact Calabi-Yau 3-fold. 
 Following the standard construction in the literature \cite{Ibanez:2012zz,Blumenhagen:2005mu,Blumenhagen:2006ci,Marchesano:2007de}, we take the orientifold action to be given by $\Omega_p (-1)^{F_L}\CR$, where $\Omega_p$ is the worldsheet parity reversal operator, ${F_L}$ is the space-time fermion number for the left-movers, and $\CR$ is an internal anti-holomorphic involution of the Calabi-Yau. This involution acts on the K\"ahler 2-form $J$ and the holomorphic 3-form $\Omega$ of $\CM$ as
\begin{equation}
\CR J=-J\ , \qquad \CR\Omega = \overline{\Omega}\label{iiaori}
\end{equation}
The fixed locus $\Pi_{\rm O6}$ of $\CR$ is given by one or several 3-cycles of $\CM$, in which O6-planes are located. In order to cancel the RR charge of such O6-planes one may introduce 4d space-time filling D6-branes wrapping three-cycles $\Pi_\alpha$ of $\CM_6$,\footnote{One may additionally consider D8-branes on coisotropic cycles as in \cite{Font:2006na}, but for simplicity we will restrict our discussion to models where only D6-branes are present.} such that the orientifold symmetry is preserved. In the absence of NS background fluxes, RR tadpole cancellation requires that the following equation in $H_3(\CM_6,\IZ)$ is satisfied
\be
\sum_\a \left([\Pi_\a] + [\CR \Pi_\a]\right) - 4 [\Pi_{\rm O6}]\, =\, 0
\label{RRtadpole}
\ee
where $\Pi_{\rm O6}$ stands for the O6-plane loci. Finally, for such D6-branes to minimise their energy and preserve the 4d $\CN=1$ supersymmetry of this background they need to wrap special Lagrangian  three-cycles. That is they need to satisfy the geometric conditions 
\begin{equation}
J|_{\Pi_\alpha}=0\qquad  {\rm and} \qquad \im\, \Omega|_{\Pi_\alpha}=0 \quad \quad \quad \forall \, a
\end{equation}
as well as to have vanishing worldvolume flux $\CF$, defined as
\begin{equation}\label{Eq:DbraneWVFlux}
\CF = B|_{\Pi_\alpha} - \sigma F, 
 \end{equation}
 with $\sigma = \frac{l_s^2}{2\pi}$ and the string length given by $l_s = 2\pi \sqrt{\a'}$.

In the absence of D-brane moduli, the 4d effective action for the closed string sector of these constructions has been analysed in great detail, see for instance \cite{Grimm:2004ua,Grimm:2011dx,Kerstan:2011dy}. In particular, the moduli space of closed string deformations and its related K\"ahler potential can be described as follows. On the one hand there are $h^{1,1}_-(\CM_6)$ complexified K\"ahler moduli defined as
\be \label{Eq:JcExpansion}
J_c\, =\, B + ie^{\phi/2} J\, =\, T^a  \omega_a
\ee
where $\phi$ is the 10d dilaton, $J$ is computed in the Einstein frame and $l_s^{-2} \omega_a$ are harmonic representatives of $H^2_-(\CM_6, \IZ)$ that can be defined as $\om_a = d \im J_c/d \im T^a$. At large volumes compared to the string scale the tree-level K\"ahler potential for these moduli is given by
\be
K_K \,  = \, -{\rm log} \left(\frac{i}{6} \CK_{abc} (T^a - \bar{T}^a)(T^b - \bar{T}^b)(T^c - \bar{T}^c) \right) 
\label{KK}
\ee
with $\CK_{abc} = l_s^{-6} \int_{\CM_6} \om_a \wedge \om_b \wedge \om_c \in \IZ$ the triple intersection numbers of the compactification manifold. At this level the B-field axions $b^a \equiv \re \, T^a$ display a continuous shift symmetry only broken by worldsheet instantons and $e^{-K_K}$ is a cubic polynomial on the moduli $t^a \equiv \im\, T^a$.

On the other hand, the moduli space of complex structure deformations is encoded in terms of the harmonic three-form 
\be
\Omega_c = C_3 + i \re (C \Omega) \quad  \in \quad \CH^3_+(\CM_6)
\ee
which is even under the orientifold involution. Here $C_3$ is the three-form RR potential and $C \equiv e^{-\phi} e^{\oh(K_{CS}-K_{K})}$ stands for a compensator term with $K_{CS} = - {\rm log}\left( \frac{i}{l_s^6} \int \Omega \wedge \bar{\Omega}\right)$. In order to translate this quantity into 4d chiral fields one first describes $\Omega$ in terms of a symplectic integer basis $(\a_\lam,\b^\lam) \in H^3 (\CM_6, \IZ)$ as $\Om = X^\lam \a_\lam - \CF_\lam\b^\lam$ \cite{Candelas:1990pi}. Then one uses the orientifold action to split such a basis into even $(\a_K, \b^\Lam) \in H_+^3(\CM_6)$ and odd $(\a_\Lam, \b^K) \in H_-^3(\CM_6)$ three-forms and defines the chiral fields 
\be
N'^{K}\, =\, l_s^{-3} \int_{\CM_6} \Omega_c \wedge \b^K \qquad \qquad U'_{\Lam}\, =\, l_s^{-3} \int_{\CM_6} \Omega_c \wedge \a_\Lam
\label{defNU}
\ee
which have a well-defined counterpart in mirror type IIB orientifolds \cite{Grimm:2004ua}. Finally, one imposes the orientifold constraints to obtain the K\"ahler potential
\be
K_Q\,  = \, -2\,{\rm log} \left(\frac{1}{4} \left[ \re(C\CF_\Lam)\im(CX^\Lam) - \re(CX^K)\im(C\CF_K) \right]  \right)
\label{KQ}
\ee
which should then be translated into the variables $n'^{K} \equiv \im \, N'^{K}$ and $u'_\Lam \equiv \im\, U'_\Lam$ on which it depends. As pointed out in \cite{Kerstan:2011dy} one can always perform a symplectic transformation so that all moduli are of the kind $N'^K$. In this case showing such dependence is relatively easy, as we have
\be
K_Q\,  = \, -2\,{\rm log} \left(-\frac{1}{4} \im(\CF_{KL})\, n'^K n'^L \right)
\label{KQN}
\ee
where $\im\, \CF_{KL}$ is a homogeneous function of zero degree on the $n'^K$. In general one can show that $e^{-K_Q/2}$ is a homogeneous function of degree two on $n'^K$ and $u'_\Lam$, as we discuss in Appendix \ref{ap:details}.

This effective field theory becomes more involved as soon as we introduce open string degrees of freedom \cite{Grimm:2011dx,Kerstan:2011dy,Camara:2011jg,Marchesano:2014bia}. In particular it was found in \cite{Grimm:2011dx,Kerstan:2011dy} that the complex structure moduli are redefined in the presence of D6-brane moduli. Rewriting the K\"ahler potential (\ref{KQ}) in terms of the new 4d chiral fields modifies its expression and introduces a dependence in the open string modes. In the following sections we would like to analyse such modifications, how they affect the K\"ahler potential for closed and open string fields and their implications for the scalar potential governing both. As we will see, a key ingredient of our analysis will be the periodic directions that appear in open string moduli space, and the discrete shift symmetries that they correspond to. 

Given a particular compactification, one may describe a point in open string moduli space by considering a set of special Lagrangian three-cycles $\{\Pi_\a^0\}$ that satisfy the RR tadpole condition (\ref{RRtadpole}) and where D6-branes are wrapped. One may now move in this moduli space by deforming the reference three-cycle $\Pi_\a^0$ to a homotopic special Lagrangian three-cycle $\Pi_\a$. If the deformation is infinitesimal we can describe it in terms of a normal vector $X$, and define the open string moduli of a D6-brane wrapping such a three-cycle as
\be
\Phi_\a^i  \, =\,  \frac{2}{l_s^4} \int_{\Pi_\a^0} \left(\sig A - \iota_{X} J_c \right) \wedge \rho^i \, =\, \theta_\a^i - T^a (\eta_{\a\, a}^0)^i{}_j   \varphi^j _\a
\label{defPhi}
\ee
where $A$ is the D6-brane gauge potential along $\Pi_3^0$ containing the Wilson lines degrees of freedom,\footnote{The normalisation for $\Phi$ is such that the periodicity $\int_{\g} A \sim \int_{\g} A + \pi$ for D-branes on top of orientifold planes is translated into $\theta \sim \theta +1$. A similar statement holds for the position moduli.} and $l_s^{-2} \rho^i \in \CH^2(\Pi_\a^0, \IZ)$ is a basis of quantised harmonic two-forms in $\Pi_\a^0$.\footnote{By taking $\rho^i$ harmonic we are selecting the lowest of a tower of Kaluza-Klein open string modes, but one may extend this definition to the full tower by taking an appropriate basis of quantised two-forms \cite{Grimm:2011dx}. For most pourposes we will assume a truncation to the lightest states of each open string KK tower.} Here $X = \oh l_s X_j\varphi^j$ is a linear combination of normal vectors to $\Pi_\a^0$ preserving the special Lagrangian condition, that is such that $[\CL_{X_j} J]_{\Pi_\a^0} = [\CL_{X_j} \Om]_{\Pi_\a^0} = 0$ with $\CL_{X_j}$ the Lie derivative along $X_j$. Following \cite{McLean} this implies that $\iota_{X_j} J|_{\Pi_\a^0}$ is proportional to a harmonic one-form on $\Pi_\a^0$,\footnote{Notice that the same is not true for $\iota_{X_j}J_c|_{\Pi_\a^0}$, whose real part could be a non-harmonic one-form. However, one may always choose the profile for $A$ such that $A - \iota_{X} J_c|_{\Pi_\a^0}$ is harmonic. Alternatively one may take $\rho^i$ to be harmonic two-forms, as we do here.}  and therefore $i,j = 1, \dots, b_1(\Pi_\a^0)$. Finally we define
\be
(\eta_{\a\, a}^0)^i{}_j\, \equiv\, l_s^{-3} \int_{\Pi_\a^0} \iota_{X_j} \omega_a \wedge \rho^i 
\label{etas}
\ee
to make manifest the implicit dependence of $\Phi$ on the K\"ahler moduli. 

For finite homotopic deformations $\Pi_\a = {\rm exp}_{X}(\Pi_\a^0)$ the dependence on the deformation parameters $\varphi^j$ should be computed by a normal coordinate expansion \cite{Grimm:2011dx,Kerstan:2011dy}, which adds to the above linear behaviour a higher order dependence on the $\varphi$'s. The open string moduli can then be expressed as
\be
\Phi_\a^i  \, =\,  \theta_\a^i - T^a f_{\a\, a}^i
\label{defPhif}
\ee
where the functions $f_{\a\, a}^i(\varphi)$ satisfy the differential equation
\be
\frac{\p f_{\a\, a}^i}{\p \varphi^j}\, =\,  (\eta_{\a\, a})^i{}_j\, \equiv\, l_s^{-3} \int_{\Pi_\a} \iota_{X_j} \omega_a \wedge \rho^i
\label{deff}
\ee
so that by imposing $f_{\a\, a}^i(\varphi^j = 0) = 0$ we recover (\ref{defPhi}) from the leading term in the Taylor expansion $f_{\a\, a}^i = (\eta_{\a\, a}^0)^i{}_j   \varphi^j _\a +\dots$ In general $(\eta_{\a\, a})^i{}_j$ and so $f_{\a\, a}^i$ may further depend on the Calabi-Yau metric, and therefore on the closed string moduli $\im\, T^a$, $\im\, N'^K$, $\im\, U'_\Lam$. 

In simple compactifications like toroidal orientifolds, the $\eta_a$'s are independent of $\varphi$, and so the definition (\ref{defPhi}) is exact. Moreover, the normal vectors $X_j$ can be chosen such that the $\eta_a$'s are integer numbers and both $\theta$'s and $\varphi$'s are periodic variables of unit period \cite{Cremades:2003qj,Berg:2011ij}. This last statement will remain true for the $\theta$'s in general Calabi-Yau compactification while for the $f_a$'s things may become more complicated. The best way to analyse their periodicity is to define the open string moduli in terms of integration chains, along the lines of \cite{Hitchin:1997ti,Kachru:2000an,Grimm:2011dx}.  

When the three-cycle $\Pi_\a^0$ is homotopically deformed to $\Pi_\a$, a one-cycle $\g_i^0$ in the Poincar\'e dual class to $\rho^i$ will sweep a two-chain $\G_{\a}^i$ in $\CM_6$, such that $\p\G^i_\a = \g_i - \g_i^0$ with $\g_i$ the corresponding one-cycle in $\Pi_\a$. One can the define the complexified open string coordinates as \cite{Hitchin:1997ti,Kachru:2000an}
\be
\Phi^i_\a  \, =\,  \frac{2}{l_s^{2}} \int_{\G_{\a}^i} \sig \tilde{F} - J_c
\label{defPhich}
\ee
where $\tilde F$ is an extension of the worldvolume field strength $F=dA$ to the two-chain $\G_\a^i$ such that $\int_{\G_2^i} \tilde{F}\, =\, \int_{\g_i} A - \int_{\g_i^0} A$. Using Lefschetz duality one may then rewrite this as
\be
\Phi^i_\a  \, =\,  \frac{2}{l_s^{4}} \int_{\cc_4^\a} \left( \sig \tilde{F} - J_c \right) \wedge \tilde{\rho}^i
\label{defPhi4ch}
\ee
as done in \cite{Grimm:2011dx}. Here $\cc_4^\a$ is the four-chain swept by the three-cycle $\Pi_\a$ and $\tilde{\rho}^i$ the quantised two-form on $\cc_4^\a$ that pulls-back to $\rho^i$ on its boundary, while $\tilde F$ is now extended to the whole of $\cc_4^\a$. By comparing these definitions to (\ref{defPhif}) one obtains that
\be
f_{\a \, a}^i\, =\, \frac{2}{l_s^{2}} \int_{\G_{\a}^i} \om_a\, =\,  \frac{2}{l_s^{4}} \int_{\cc_4^\a}\om_a \wedge \tilde{\rho}^i
\label{deffch}
\ee
which clearly satisfies (\ref{deff}). From this perspective it is easy to understand when the functions $f_a$ describe periodic coordinates in the D6-brane moduli space. If a homotopic special Lagrangian deformation is such that $\cc_4^\a$ is a four-cycle in $\CM_6$, the D6-brane system has returned to its original position after performing a loop in its moduli space. This should correspond to a discrete symmetry of the theory, just like shifts of Wilson lines by their period. Notice that if $\cc_4^\a$ is a four-cycle then the $f_a$'s are integer numbers,\footnote{In fact they are even integer numbers. However we will also consider the possibility where $\cc_4^\a\cup\CR\cc_4^\a$ is a four-cycle without $\cc_4^\a$ being so, which implies odd $f_a$'s.} and so the shifts that are generated by periodic directions in the D6-brane moduli space are
\be
\Phi_\a^i\, \raw\, \Phi_\a^i + k_\a^i \qquad \text{and} \qquad \Phi_\a^i\, \raw\, \Phi_\a^i - T^a r_{\a\, a}^i
\label{shiftPhi}
\ee
with $k_\a^i,  r_{\a\, a}^i \in \IZ$. As we will see later on, such discrete shifts are directly related to the discrete gauge symmetries and the multi-branched structure of F-term scalar potentials in type II compactifications with fluxes and D-branes.

Another set of quantities that will become important in the formulation of the K\"ahler potential are the functions $\{g_{\a\, i}^{K}, g_{\a\, \Lam\, i}\}$ on the deformation parameters $\varphi$'s. Such functions provide an alternative parameterisations of the D6-brane moduli space and are defined by the differential equations  \cite{Hitchin:1997ti}
\be
\frac{\p g_{\a\, i}^{K}}{\p \varphi^j}\, =\,  (\CQ^K_\a)_{ij} \qquad \text{and} \qquad \frac{\p g_{\a\, \Lam\, i}}{\p \varphi^j}\, =\, (\CQ_{\a\, \Lam})_{ij}
\label{pgQ}
\ee
where 
\be
(\CQ^K_\a)_{ij} \, =\, l_s^{-3} \int_{\Pi_\a} \iota_{X^j} \beta^K \wedge \zeta_i \qquad \qquad (\CQ_{\a\, \Lam})_{ij}\, =\, l_s^{-3} \int_{\Pi_\a} \iota_{X^j} \a_\Lam \wedge \zeta_i
\label{Qs}
\ee
Here $\Pi_\a = \text{exp}_X (\Pi_\a^0)$ is again a homotopic special Lagrangian deformation of $\Pi_\a^0$, and $\zeta_i$ is a basis of quantised harmonic one-forms on it such that $\int_{\Pi_\alpha^0} \zeta_i \wedge \rho^j = l_s^3 \d^j_i$. Again, these functions can be expressed in terms of chain integrals as \cite{Grimm:2011dx}
\be
\frac{2}{l_s^{3}} \int_{\Sig_\a^i} \im\, (C \Om)\, =\, \frac{2}{l_s^{4}}\int_{\cc_4^\a} \im\, (C \Om) \wedge \tilde{\zeta}_i\, =\, - g_{\a\, i}^{K} \im(C\CF_K) + g_{\a\, \Lam\, i} \im(CX^\Lam)
\ee
where $\Sig_\a^i$ is the three-chain swept by the two-cycle Poincar\'e dual to $\zeta_i$, and $\tilde \zeta_i$ is the extension of this one-form to the four-chain $\cc_4^\a$. More explicitly we have that
\be
g_{\a\, i}^{K} \, =\, \frac{2}{l_s^{4}} \int_{\cc_4^\a}  \b^K \wedge \tilde{\zeta}_i \qquad \text{and} \qquad g_{\a\, \Lam\, i} \, =\, \frac{2}{l_s^{4}} \int_{\cc_4^\a}  \a_\Lam \wedge \tilde{\zeta}_i .
\label{defgs}
\ee

\section{The scalar potential from Minkowski three-forms}
\label{s:3forms}

The space of background and D-brane deformations described in the last section will be subject to a scalar potential in certain type IIA compactifications. In particular, both K\"ahler and complex structure moduli will develop an F-term scalar potential when NSNS and RR background fluxes are present \cite{Louis:2002ny,Grimm:2004ua}. In the absence of open string deformations, this potential can been reproduced by applying the usual 4d supergravity expression or by direct dimensional reduction. The latter method involves integrating out the degrees of freedom associated to three-form fields in $\IR^{1,3}$ which give a non-vanishing contribution to the potential \cite{Louis:2002ny,Grimm:2004ua}. In fact, as shown in \cite{Bielleman:2015ina} one may describe the full F-term scalar potential purely in terms of contributions coming from Minkowski three-forms if one performs the dimensional reduction in the democratic formulation of type IIA supergravity. In the following we will adopt this latter approach, as it allows to incorporate the D-brane moduli into the computation and derive a scalar potential for open and closed string modes simultaneously \cite{Grimm:2011dx,Kerstan:2011dy}. 

For simplicity let us consider a Calabi-Yau orientifold compactification where only RR background fluxes are present, ignoring for now the presence of localised sources. Working in the democratic formulation we have the RR $p$-form potentials $C_p$ with $p=1,3,5,7,9$ that can be arranged in the polyforms
\be
{\bf C} \, =\, C_1 + C_3 + C_5 + C_7 + C_9  \quad \quad {\rm or} \quad \quad {\bf A} \, =\, {\bf C} \wedge e^{-B}
\ee
respectively dubbed {C} and {A}-basis for the RR potentials in \cite{Bergshoeff:2001pv}. The corresponding gauge invariant field strengths are given by \cite{Bergshoeff:2001pv,Villadoro:2005cu}
\be
{\bf G} \,=\, d{\bf C} - H \wedge {\bf C} + {\bf \bar{G}} \wedge e^B\, =\, \left(d{\bf A} + {\bf \bar{G}}\right) \wedge e^B
\label{bfG}
\ee
%
with ${\bf \bar{G}}$ a formal sum of harmonic $(p+1)$-forms of $\CM_6$ to be thought as the background value for the internal RR fluxes defined in the A-basis. This basis is particularly adequate to apply Dirac quantisation, since a D$(p-1)$-brane will couple to the potential $A_{p}$ via its Chern-Simons action, and so the standard reasoning gives the quantisation condition \cite{Marolf:2000cb}
\be
\frac{1}{l_s^{p}} \int_{\pi_{p+1}} dA_{p} + \bar{G}_{p+1} \, \in \, \IZ
\label{Diracq}
\ee
for any cycle $\pi_{p+1}$ in the internal space not intersecting a localised source like a background D-brane. In the A-basis a source wrapping a cycle $\Pi_a$ enters the Bianchi identity as
\be
d\left(e^{-B}\wedge {\bf G}\right) \, =\, d\left(d{\bf A} +  {\bf \bar{G}}\right)\, =\, \sum_a \d(\Pi_a) \wedge e^{-\sig F_a}
\label{BIG}
\ee
with $\d(\Pi_a)$ the delta-function with support on $\Pi_a$ and indices transverse to it, while $F_a$ is the quantised worldvolume flux threading $\Pi_a$. In the absence of localised sources the rhs of (\ref{BIG}) vanishes and the $A_p$ are globally well-defined, so they do not contribute to the integral in (\ref{Diracq}) which becomes a quantisation condition for the fluxes $ \bar{G}_{p+1}$. We can then define flux quanta in terms of the integer cohomology of $\CM_6$, namely as
\begin{equation}
m \, = \, l_s \bar{G}_0 ,  \qquad  m^a\, =\, \frac{1}{l_s} \int_{\tilde \pi^a} \bar{G}_2 , \qquad  e_a\, =\, \frac{1}{l_s^3} \int_{\pi_a} \bar{G}_4 , \qquad e_0 \, =\, \frac{1}{l_s^5} \int_{{\cal M}_6} \bar{G}_6 
\label{RRfluxes}
\end{equation}
where $\tilde \pi^a \in H_2^-({\cal M}_6,\Z)$ and $\pi_a\in H_4^+ ({\cal M}_6, \Z)$. These definitions need to be generalised if we take into account the effect of localised sources in the above Bianchi identity, as we discuss in the next section. 
Finally, as we have doubled the $p$-form degrees of freedom we need to impose the Hodge duality relations
\begin{equation}
G_{2n} = (-)^{n} \star_{10}  G_{10-2n}.
\end{equation}
which can be done either by hand or by adding a series of Lagrange multipliers to the 10d supergravity action, as done in Appendix \ref{ap:dimred}.
 
 To proceed we define a set of Minkowski four-form field strengths arising from the dimensional reduction of the 10d RR field strengths
  \be
 G_4 = F_4^0 + \ldots \quad G_6 =  F_4^a \wedge \omega_a + \dots \quad G_8 = \tilde F_{4\, a} \wedge \tilde\omega^a +\dots \quad G_{10} = \tilde F_{4} \wedge \om_6 + \dots
 \label{F4form}
 \ee
 Here the four-forms $(F_4^0, F_4^a, \tilde F_{4\, a}, \tilde F_4)$ have their indices in $\IR^{1,3}$ and $(\om_a, \tilde \om^a, \om_6)$ are harmonic forms of $\CM_6$.\footnote{In terms of the dimensional reduction of the RR potential to Minkowski three-forms
 \be
 C_3 = c_3^0 + \ldots \quad C_5 =  c_3^a \wedge \omega_a + \dots \quad C_7 = \tilde d_{3\, a} \wedge \tilde\omega^a +\dots \quad C_9 = \tilde d_{3} \wedge \om_6 + \dots
 \label{A3form}
 \ee
 We have that
\be
F_4^0\, =\, dc_3^0 \qquad F_4^a\, =\, dc_3^a -db^a \wedge c_3^0 \qquad \tilde F_{4\, a} \, =\, d\tilde d_{3\, a} -\CK_{abc}db^b \wedge c_3^c\qquad \tilde F_4\, =\, d\tilde d_3 - db^a \wedge \tilde d_{3\, a}
\label{M4forms}
\ee}
 In particular $l_s^{-2} \omega_a$ are the harmonic representatives of $H^2_-(\CM_6, \IZ)$ defined in the last section and $l_s^{-4}\tilde \omega^a$ is a dual basis of harmonic forms in $H_+^{4}({\cal M}_6, \Z)$ in the sense that
 \begin{equation}
\int_{{\cal M}_6} \omega_a \wedge \tilde \omega^b = l_s^6 \delta_a^b, \qquad a,b, \in\{1, \ldots , h^{1,1}_-\}.
\end{equation} 
Finally $l_s^{-6} \om_6 =  d{\rm vol}_{\CM_6}/ \sqrt{g_{\CM_6}}$ represents the unique harmonic six-form with unit integral over $\CM_6$.
Following \cite{Bielleman:2015ina}, such Minkowski four-forms enter the 4d effective action as
\bea\nonumber
& &  - \oh \frac{e^{-K}}{32\kappa_4^2} \int_{\IR^{1,3}} \frac{1}{4} F_4^0 \wedge * F_4^0 + g_{ab} F_4^a \wedge * F_4^b + \frac{e^{-3\phi}g^{ab}}{16\hat{V}_6^2} \tilde F_{4\, a} \wedge * \tilde F_{4\, b} + \frac{e^{-3\phi}}{4\hat{V}_6^2} \tilde F_4 \wedge * \tilde F_4
 \\
& & + \frac{1}{4\kappa_4^2} \int_{\IR^{1,3}} F_4^0 \rho_0 + F_4^a \rho_a + \tilde F_{4, a} \tilde \rho^a + \tilde F_4 \tilde \rho 
\label{S4form}
\eea
where $\hat{V}_6 = l_s^{-6} {\rm Vol} (\CM_6)$ stands for the covering space compactification volume in the Einstein frame and in string units. In deriving the above expression we have performed the 4d Weyl rescaling  $g_{\mu \nu}\rightarrow\frac{g_{\mu \nu}}{ \hat V_6/2}$, we have used that
\be
e^{K}\, =\, \frac{e^{-\phi/2}}{8 \hat{V}_6^3}
\label{eK}
\ee
and that 
\begin{equation}
g_{ab} = \frac{e^{-\phi}}{4 \hat{V}_6 l_s^6 } \int_{{\cal M}_6} \omega_a \wedge \star_6 \omega_b  \qquad g^{ab} = \frac{4 \hat{V}_6 e^\phi}{l_s^6} \int_{{\cal M}_6} \tilde \omega^a \wedge \star_6 \tilde \omega^b
\end{equation}
 represent the usual metric for harmonic two-forms and its inverse. Finally we have that
 \be
 \begin{array}{rcl}
l_s \rho_0 & = &  e_0 + b^a e_a + \oh \CK_{abc}m^ab^bb^c + \frac{m}{6} \CK_{abc}b^ab^bb^c \\
- l_s \rho_a & = & e_a +\CK_{abc}m^bb^c + \frac{m}{2} \CK_{abc}b^bb^c\\
l_s \tilde \rho^a & = & m^a + mb^a\\
- l_s \tilde \rho & = & m
 \end{array}
 \label{rhos}
 \ee

We may now integrate out the Minkowski four-forms from (\ref{S4form}), obtaining the following scalar potential
\be
V_{\rm RR}\, =\,  \frac{1}{\kappa_4^2} e^K \left[4 \rho_0^2 + g^{ab} \rho_a \rho_b + 16 e^{3\phi} \hat{V}_6^2 g_{ab} \tilde \rho^a \tilde \rho^b + 4 e^{3\phi} \hat{V}_6^2 \tilde \rho^2 \right]
\label{VRR}
\ee
which is nothing but the usual type IIA RR flux potential  \cite{Louis:2002ny}
 \be 
  \label{expotRR}
 \begin{array}{rcl}
l_s^2 \kappa_4^2 V_{RR} (e_0, e_a, m^a, m) & = & e^K 4 \left( e_0 + b^a e_a + \oh \CK_{abc}m^ab^bb^c + \frac{m}{6} \CK_{abc}b^ab^bb^c\right)^2\\
 & + & e^K  g^{ab} \left( e_a +\CK_{acd}m^cb^d + \frac{m}{2} \CK_{acd}b^cb^d\right) \left( e_b +\CK_{bef}m^eb^f + \frac{m}{2} \CK_{bef}b^eb^f\right) \\
 & + & \frac{4}{9} e^K  \CK^2 g_{ab} \left( m^a + mb^a \right)  \left( m^b + mb^b \right)  \, +\,  \frac{1}{9} e^K \CK^2 m^2
 \end{array}
 \ee 
 where we have used that $\CK \equiv \CK_{abc} t^a t^b t^c = 6  e^{3\phi/2}  \hat{V}_6$. 
 Alternatively, one may deduce this scalar potential from the dimensional reduction procedure described in Appendix \ref{ap:dimred}.
 
\subsubsection*{Adding a single D6-brane}
 
 This method to obtain the scalar potential has the advantage that it allows to incorporate the open string scalars in a very straightforward way. Indeed, one just needs to add to the 4d action (\ref{S4form}) the extra four-forms couplings that arise form the dimensional reduction of the D-brane Chern-Simons actions, and then proceed as before. As in \cite{Grimm:2011dx,Kerstan:2011dy}, let us first consider the effect of a single D6-brane wrapping a three-cycle $\Pi_\a$ which is a homotopic special Lagrangian deformation from the reference cycle $\Pi_\a^0$. Furthermore although it may break supersymmetry, we also consider the following worldvolume flux threading the three-cycle
\be
\sig F\, =\, \sig d A + n_{F\, i}\, \rho^i \qquad n_{F\, i} \in \IZ
\ee
 where the harmonic two-forms $\rho^i$ have been defined in (\ref{defPhi}). The CS action reads
\bea
\nonumber
S_{\rm CS}^{\rm D6} & = & \mu_6 \int_{\IR^{1,3} \times \Pi_\a^0} e^{\CL_{X}}  A_7 + \sig F \wedge e^{\CL_{X}} A_5 + \oh \sig^2 F^2 \wedge e^{\CL_{X}}A_3 + \dots\\
& = & \mu_6 \int_{\IR^{1,3} \times \Pi_\a^0} {\rm exp}_{X} ({\bf A}) \wedge e^{\sig F} \, =\,  \mu_6 \int_{\IR^{1,3} \times \cc_4^\a}  \hspace*{-.5cm} d {\bf A} \wedge e^{\sig \tilde F}
\label{CSD6}
\eea
 where $\mu_6 = 2\pi/l_s^7$. As in the previous section $X = \oh l_s \varphi^j X_j$ parametrises the homotopic special Lagrangian deformation $\Pi_\a = {\rm exp}_X(\Pi_a^0)$, $\cc_4^\a$ is the corresponding four-chain such that $\p\cc_4^\a = \Pi_\a - \Pi_\a^0$ and $\tilde F$ is the extension of the field strength made in (\ref{defPhich}). Expanding the RR potentials as in (\ref{A3form}), we find the following couplings of the D6-brane scalars to the Minkowski four-forms arising from the bulk
 \be
 \frac{1}{4\kappa_4^2} \int_{\IR^{1,3}} F_4^0\, \upsilon_0 + F_4^a \upsilon_a + \tilde F_{4\, a}\, \tilde \upsilon^a\, .
 \label{S4formop}
 \ee
Here we have defined
  \be
 \begin{array}{rcl}
l_s \upsilon_0 & = &  n_{F\, i}\th^i - b^a n_{F\, i} f_a^i - b^a n_{a\, i} \th^i + n_{a\, i} f_c^i  b^a b^c  \\
l_s \upsilon_a & = & n_{a\, i}\theta^i - n_{a\, i} b^c f_c^i  + n_{F\, i} f_a^i - b^c n_{c\, i} f_a^i \\
l_s \tilde \upsilon^a & = &  q^a
 \end{array}
 \label{varrhos}
 \ee
with $f_a^i$ as in (\ref{deffch}) and 
 \bea
 n_{a\, i} & = & \frac{1}{l_s^{3}} \int_{\Pi_\a} \om_a \wedge \zeta_i \\
 q^a & = & \frac{2}{l_s^{4}} \int_{\cc_4^\a} \tilde \om^a
 \eea
with the harmonic one-forms $\zeta_i$ defined as in (\ref{Qs}). By using that the pull-back and the wedge product commute, one can show that these quantities satisfy
\begin{equation}
n_{a\, i} \frac{\p f_b^i}{\p{\varphi^j}}  + n_{b\, i}  \frac{\p f_a^i}{\p{\varphi^j}} = {\cal K}_{abc}  \frac{\p q^c}{\p{\varphi^j}} \qquad \Raw \qquad n_{a\, i} f_b^i  + n_{b\, i}  f_a^i = {\cal K}_{abc}  q^c
\label{pullbackid}
\end{equation}
 which can be used to simplify (\ref{varrhos}). 
 
 Incorporating the couplings (\ref{S4formop}) into the 4d effective action and then integrating the Minkowski four-forms will result into performing the following replacements in the expression (\ref{VRR}) for the scalar potential
\be
\rho_0 \, \mapsto \, \varrho_0 = \rho_0 + \upsilon_0 \qquad \rho_a \, \mapsto \, \varrho_a = \rho_a - \upsilon_a \qquad \tilde \rho^a \, \mapsto \, \tilde \varrho^a = \tilde \rho^a + \tilde \upsilon^a
\ee
while leaving $\tilde \rho$ invariant. Equivalently, one may replace the flux quanta that enter (\ref{expotRR}) by the following quantities
\be
\tilde e_0 = e_0  + n_{F\, i}  \theta^i \qquad \tilde e_a = e_a - n_{a \, i}\th^i - n_{F\, i} f_a^i \qquad \tilde m^a =  m^a + q^a \label{shiftedflux}
\ee
while leaving the Romans mass untouched. The scalar potential that we then obtain is 
\be
V_{\rm sc}\,=\, V_{\rm RR + CS} + V_{\rm DBI}
\label{Vsc}
\ee
where $V_{\rm DBI}$ stands for the total tension of the D6-branes minus that of the O6-planes, while
\bea\nonumber
V_{\rm RR + CS}  & = & \frac{1}{\kappa_4^2} e^K \left[4 \varrho_0^2 + g^{ab} \varrho_a \varrho_b + \frac{4}{9} e^K  \CK^2 g_{ab} \tilde \varrho^a \tilde \varrho^b + \frac{1}{9} e^K \CK^2 \tilde \rho^2 \right]\\ & =  & V_{\rm RR} (\tilde e_0, \tilde e_a, \tilde m^a, m)
\label{VRRCS}
\eea
with $V_{\rm RR}$ given by (\ref{expotRR}) and $\CK$ defined below it. For the explicit expression of $V_{\rm RR + CS}$ in terms of closed and open string fields see eq.(\ref{expotRROpen}).

The scalar potential (\ref{VRRCS}) includes both closed and open string deformations, and reproduces the results found in \cite{Grimm:2011dx,Kerstan:2011dy}. Particularly important for our purposes it the  series of discrete shift symmetries that it contains both for closed and open string axions. The symmetries of the closed string sector are already present in scalar potential (\ref{VRR}) and correspond to simultaneous discrete shifts in the B-field axions and backgrounds fluxes, as described in detail in \cite{Bielleman:2015ina}. In this sense, the open-closed scalar potential (\ref{Vsc}) adds new discrete shift symmetries related to the D6-brane Wilson line and position deformations. Indeed, on the one hand by definition $V_{\rm DBI}$ does not depend on the Wilson lines. On the other hand, $V_{\rm RR + CS}$ only depends on them through the shifted fluxes (\ref{shiftedflux}), dubbed {\it dressed} fluxes henceforth. The latter are left invariant by the discrete shifts
\be
\theta^i \, \raw \, \theta^i + k^i \qquad e_0 \, \raw \, e_0 - k^i n_{F\, i} \qquad e_a \, \raw \, e_a + k^i n_{a\, i} 
\label{shiftWL}
\ee
with $k^i$ such that $k^in_{F\, i}, k^i n_{a\, i} \in \IZ$. Consequently the full scalar potential is left invariant by these shifts. Because $n_{F\, i}, n_{a\, i} \in \IZ$, a particular solution is given by taking $k^i \in \IZ$, which corresponds to the first kind of shift in (\ref{shiftPhi}). Note that such discrete symmetries are nothing but large gauge transformations related to the D6-brane Wilson lines, and therefore discrete gauge symmetries. As such they should be present not only in the scalar potential but also at the level of the superpotential, as we will discuss in the next section.

Similarly, if there are periodic directions in the moduli space of D6-brane positions we may formulate the following discrete shifts
\be
f_a^i \, \raw\,  f_a^i + r_a^i \qquad e_a \, \raw \, e_a + n_{F\, i} r_a^i \qquad m^a\CK_{abc} \, \raw \, m^a\CK_{abc} - [n_{b\, i}r^i_{c} + n_{c\, i}r^i_{b}]
\label{shiftpos}
\ee
where now $r_a^i$ is such that $n_{F\, i} r_a^i \in \IZ$ and $n_{(b\, i}r^i_{c)} = s^a\CK_{abc}$ with $s^a \in \IZ$. These conditions are satisfied whenever the shift correspond to the D6-brane sweeping a four-cycle in $\CM_6$, that is to the second kind of shift in (\ref{shiftPhi}). Again, because the D6-brane is returning to the same position, $V_{\rm DBI}$ is left invariant under these shifts.\footnote{Namely because $V_{\rm DBI}$ only depends on the induced metric and B-field. This could change if we were considering compactifications with non-trivial NS flux $H_3$.} 

In both cases, the presence of these discrete symmetries is directly related to the multi-branched structure of the scalar potential, typical of models with axion-monodromy. The multi-branched structure of the closed string sector of this potential was analysed in \cite{Bielleman:2015ina}. In this sense, the presence of open string modes in the potential and the discrete symmetries (\ref{shiftWL}) and (\ref{shiftpos}) describe how open string modes are related to these branches of the scalar potential, as already pointed out in \cite{Escobar:2015ckf} for the case of the Wilson line. 

 \subsubsection*{Adding all the D6-branes}

Let us now consider a full compactification with several D6-branes, wrapping the special Lagrangian three-cycles $\Pi_\a$ and their orientifold images. Recall that a consistent configuration must satisfy the RR tadpole condition (\ref{RRtadpole}), which is equivalent to the existence of a four-chain $\Sigma_4 \subset \CM_6$ that connects all the D6-branes and O6-planes. Physically, one can interpret this four-chain as follows. If we wrap a D6-brane on it, we will construct a domain wall in 4d connecting two different vacua: one of them with  all the D6-branes on top of the O6-planes and the other one with the D6-branes wrapping the three-cycles $\Pi_\a$ and their orientifold images.

Now, considering such a global configuration allows to take into account terms of the Chern-Simons action which we implicitly neglected in the single D6-brane case, when computing the couplings (\ref{S4formop}). These terms are those without derivatives, namely
\be
\mu_6 \sum_\a  \int_{\IR^{1,3} \times \Pi_\a^0} \hspace*{-.5cm}  A_7 + \sig F \wedge A_5 + \dots \, =\, \frac{\mu_6}{2} \int_{\IR^{1,3} \times  \Sig_4^0}  \hspace*{-.5cm} d {\bf A} \wedge e^{\sig \tilde F}\, =\, \frac{\mu_6}{2} \int_{\IR^{1,3} \times \Sig_4^0}  \hspace*{-.5cm} ({\bf G} \wedge e^{- B} - \bar{\bf G}) \wedge e^{\sig \tilde F }
\ee
where in the first equality we have used that $A_7$ and $F$ vanish on top of the O6-planes and then applied Stokes' theorem on the reference four-chain $\Sigma_4^0$, defined such that $\p\Sigma_4^0 = \sum_\a (\Pi_\a^0 - \CR\Pi_\a^0) - 4 \Pi_{\rm O6}$. For this one needs to extend of the worldvolume flux $F$ from the boundaries to the four-chain $\Sigma_4^0$ connecting them, extension which we dubbed $\tilde F$. In terms of the D6-brane domain wall described above, $\tilde F$ would be the actual worldvolume flux of the D6-brane along $\Sigma_4^0$. 

Setting the Wilson lines on all the D6-branes to zero, the rhs of this equation gives 4d couplings of the form (\ref{S4formop}), which will eventually translate into shifts of the flux quanta. In particular we have the shifts
\be
e_0 \raw e_0 + \frac{1}{8\pi^2} \int_{\Sig_4^0} \tilde F \wedge \tilde F \qquad
e_a  \raw  e_a - \frac{1}{2\pi l_s^{2}} \int_{\Sig_4^0} \om_a \wedge \tilde{F} \qquad m^a  \raw  m^a + \frac{1}{l_s^{4}} \int_{\Sig_4^0} \tilde\om^a .
\label{shiftchain}
\ee

Let us now add the contribution of each D6-brane due to turning on the Wilson lines and deforming its embedding away from the reference cycles $\Pi_\a^0$. One then obtains
\be
\begin{array}{rcl}
\tilde e_0 & = & e_0  + \sum_\a n^\a_{F\, i}  \theta_\a^i + \frac{1}{2} \int_{\Sig_4^0} \frac{\tilde F}{2\pi} \wedge \frac{\tilde F}{2\pi} \\
\tilde e_a & = & e_a - \sum_\a [n^\a_{a \, i}\th_\a^i + n^\a_{F\, i} f_{\a\, a}^i] - l_s^{-2} \int_{\Sig_4^0} \om_a \wedge \frac{\tilde F}{2\pi}  \\
\tilde m^a & = & m^a + \sum_\a q_\a^a  + l_s^{-4} \int_{\Sig_4^0} \tilde\om^a
\end{array}
\label{shiftedfluxch0}
\ee
where the index $\a$ runs over each independent brane but not their orientifold images. 
Finally, since such contributions can be  described in terms of four-chains $\cc_4^\a$ such that $\p \cc_4^\a = \Pi_\a - \Pi_\a^0$, one may define a new global chain $\Sigma_4 = \Sig_4^0 + \sum_\a \cc_4^\a\cup\CR\cc_4^\a$ and define the dressed fluxes in (\ref{VRRCS}) in terms of it
\be
\tilde e_0 = e_0 + \frac{1}{2} \int_{\Sig_4} \frac{\tilde F}{2\pi} \wedge \frac{\tilde F}{2\pi}  , \quad \tilde e_a = e_a - l_s^{-2} \int_{\Sig_4} \om_a \wedge \frac{\tilde F}{2\pi} , \quad \tilde m^a =  m^a + l_s^{-4} \int_{\Sig_4} \tilde\om^a 
\label{shiftedfluxch}
\ee
where we have absorbed the Wilson line dependence in the definition of $\tilde F$. Notice that this reproduces and extends the result in \cite{Grimm:2011dx}, in the sense that it gives an expression for the scalar potential globally valid in the open string moduli space. In this last expression the discrete shift symmetries leaving invariant the potential are particularly transparent. Such discrete symmetries will be useful to determine the superpotential that corresponds to this F-term scalar potential, as we discuss next.

\section{Open-closed superpotential and axion monodromy}
\label{s:supo}

In the absence of D-branes, the superpotential generated by type IIA RR fluxes is \cite{Gukov:1999gr,Taylor:1999ii} 
\be
l_s {W}_K  \, =\,  e_0 + e_a T^a + \frac{1}{2} {\cal K}_{abc} m^a T^b T^c +m \frac{1}{6} {\cal K}_{abc} T^a T^b T^c 
\label{WK}
\ee
Indeed, one can check that plugging this superpotential and the K\"ahler potential (\ref{KK}) in the standard 4d supergravity formula (\ref{expotRR}) is recovered as an F-term scalar potential \cite{Grimm:2011dx}.

Adding D6-branes degrees of freedom should modify the superpotential as
\be
{W} = {W}_K + {W}_{\rm D6}
\label{Wtotal}
\ee
where $W_{\rm D6}$ contains open and closed string moduli. By the discussion of the last section we should impose that the full superpotential is invariant under the discrete gauge symmetries related to integer shifts of the Wilson lines and completing loops in the moduli space of special Lagrangians. If we just consider the effect of a single D6-brane on $\Pi_\a$, such symmetries amount to (\ref{shiftWL}) and (\ref{shiftpos}), fixing the new superpotential piece to
\be
l_s {W}_{\rm D6}(\Phi)  \, = \, \Phi^i(n_{F\, i} - n_{a\, i} T^a) + l_s W_{\rm D6}^0
\label{supoD6single}
\ee
where we have used (\ref{defPhif}) and defined $W_{\rm D6}^0$ as the superpotential at $\Pi_\a = \Pi_\a^0$. This superpotential contains a bilinear term of the form  $n_{i\, a} \Phi^i T^a$  whose microscopic origin was described in \cite{Marchesano:2014iea} and its applications to large field inflation in \cite{Escobar:2015ckf}. As pointed out in \cite{Marchesano:2014iea} one can derive such a bilinear superpotential from the general expression \cite{Thomas:2001ve,Martucci:2006ij}
\be
l_s \left[W_{\rm D6}(\Phi) -W_{\rm D6}^0\right]  \, =\, \frac{1}{l_s^4} \int_{\cc_4^\a} (\sig \tilde F - J_c)^2
\label{supoluca}
\ee
with $\cc_4^\a$ a four-chain such that $\p\cc_4^\a = \Pi_\a - \Pi_\a^0$. Taking a homotopic deformation and applying the definition (\ref{defPhi4ch}) one sees that the superpotential (\ref{supoD6single}) is also recovered. 

To consider the full set of D6-branes in the compactification let us follow \cite{Marchesano:2014iea} and set
\be
l_s {W}_{\rm D6}  \, =\, \frac{1}{2l_s^4} \int_{\Sig_4} (\sig \tilde F - J_c)^2
\label{supolucafull}
\ee
where $\Sigma_4$ is the four-chain connecting all the D6-branes to the O6-planes, as described below (\ref{RRtadpole}). Now, by splitting this four-chain as $\Sigma_4 = \Sigma_4^0 + \sum_\a \cc_4^\a\cup\CR\cc_4^\a$ we obtain the following generalisation of (\ref{supoD6single})
\bea\nonumber
l_s {W}_{\rm D6} & = & \frac{1}{l_s^4} \sum_\a \int_{\cc_4^\a} (\sig \tilde F - J_c)^2 + \frac{1}{2l_s^4} \int_{\Sig_4^0} (\sig \tilde F - J_c)^2\\
& = &  \sum_\a \Phi^i_\a  (n_{F\, i}^\a - n_{a\, i}^\a T^a) + l_s W_{\rm D6}^0
\label{supoD6full}
\eea
where $\Sig_4^0$ the four-chain connecting the reference three-cycles wrapped by the D6-branes, and as usual $\a$ only runs over half of the D6-branes, excluding orientifold images.
Notice that this superpotential also satisfies the appropriate invariance under discrete shifts. Indeed, either taking the definition of dressed fluxes in (\ref{shiftedfluxch0}) or (\ref{shiftedfluxch}) one can show that
\be
W \, = \, W_{K} (e_0, e_a, m^a, m) + W_{\rm D6} \, =\, W_{K} (\tilde e_0, \tilde e_a, \tilde m^a, m)
\label{dressedW}
\ee
which is the statement made in (\ref{VRRCS}) but now at the level of the superpotential. Therefore we find that the dependence of the superpotential on the open string deformations enters uniquely through the {\it dressed} fluxes $\{\tilde e_0, \tilde e_a, \tilde m^a\}$. The latter are to be thought as the gauge invariant quantities including flux quanta and axions that typically appear in models with F-term axion monodromy \cite{Marchesano:2014mla}. Therefore, following the philosophy in \cite{Dvali:2005ws,Dvali:2005an,Kaloper:2008fb,Kaloper:2011jz}, one expects that in the 4d scalar potential the dependence on the open string axions also only appears through $\{\tilde e_0, \tilde e_a, \tilde m^a\}$. This result is clearly true for the scalar potential (\ref{Vsc}) computed at tree-level, but the claim is that it should also hold after all kind of UV corrections have been taken into account, which is particularly important in order to build models of large field inflation. 

This dependence on $\{\tilde e_0, \tilde e_a, \tilde m^a\}$ is also directly related to the multi-branched, domain-wall connected structure of the scalar potential. From the viewpoint of Wilson line axion dependence this structure was partly discussed in \cite{Escobar:2015ckf}, which considered the case with $n_{F\, i} =0$. In that case the discrete symmetry (\ref{shiftWL}) means that a jump between branches in the Wilson line direction is made by nucleating a 4d domain wall made out of D4-branes wrapping the two-cycle in the homology class $n_{a\, i} \text{P.D.} [\tilde \om^a]$, as crossing such a domain wall will shift the appropriate internal four-form flux. When $n_{F\, i} \neq 0$ such a D4-brane is magnetised and carries an induced D2-brane charge, which implies a further shift in the internal six-form flux in the amount indicated by (\ref{shiftWL}). 

A similar statement holds whenever there are closed loops in the moduli space of special Lagrangian deformations, as shows the description of dressed fluxes in terms of chain integrals (\ref{shiftedfluxch}). Let us for simplicity take $n_{F\, i} =0$ and switch off the Wilson line moduli. Then we have that $\tilde e_0$ and $\tilde e_a$ do not depend on $\Sigma_4$ while
\be
 \tilde m^a \, =\,  m^a +  \frac{1}{l_s^4}\int_{\Sig_4} \tilde \om^a \, =\,  \frac{1}{l_s^4}  \int_{\Lam_4} \tilde \om^a +  \frac{1}{l_s^4} \int_{\Sig_4} \tilde \om^a
 \label{tmchain}
\ee
with $\Lam_4$ the appropriate choice of four-cycle in $\CM_6$.\footnote{In technical terms, we are computing $\tilde m^a$ by integrating $\tilde \om^a$ over a cycle in the relative homology group $H_4(\CM_6, \Pi_{\rm D6}, \IZ)$, with $\Pi_{\rm D6} = \cup_\a \Pi_\a \cup \Pi_{\rm O6}$. Also, in eq.(\ref{supotildem}) a piece of the open-closed superpotential is computed by integrating $J_c^2$ over a relative homology cycle. This formulation for the open-closed superpotential is analogous to the one for type IIB compactifications with D5-branes, see e.g. \cite{Lerche:2001cw,Lerche:2002ck,Lerche:2002yw,Lerche:2003hs}.} If a D6-brane closes a loop in moduli space by sweeping a non-trivial four-cycle $\Lam_{4 \, \a}$, then there will be a non-trivial change in the integral over $\Sig_4$, which can be compensated by replacing $[\Lam_4] \raw [\Lam_4] - [\Lam_{4\, \a}]$ in (\ref{tmchain}) or equivalently by shifting $m^a$ by integers. In the 4d effective theory, the latter corresponds to crossing a domain-wall made up of a D6-brane wrapping $\Lam_{4\, \a}$. Finally, when we switch on $n_{F\, i}$ the same will apply, but now such a domain wall D6-brane is magnetised internally and it will have induced D4-brane charge, that shift the $e_a$ quanta as well. 

As a final remark\footnote{See also the discussion in section 3.3 of \cite{Koerber:2007xk}.} let us point out that the open-closed superpotential can be described in the compact form
\be
W\, =\, \frac{1}{l_s^{6}} \int_{\CM_6} {\bf G} \wedge e^{iJ}  \, =\, \frac{1}{l_s^{6}} \int_{\CM_6} (d{\bf A} + \bar{\bf G}) \wedge e^{J_c} 
\ee
with ${\bf G}$ defined in (\ref{bfG}). In the absence of D6-branes this is obvious, since then $d{\bf A}$ is exact and does not contribute to the integral, so we recover (\ref{WK}) directly from the definitions (\ref{RRfluxes}). When we include D6-branes the flux polyform $d{\bf  A} + {\bf G}$ is still quantised, but it is not closed as it satisfies the Bianchi identity (\ref{BIG}).\footnote{Or in other words $d{\bf  A} + {\bf G}$ does no longer belong to the standard cohomology $H^*(\CM_6, \IZ)$ but it does belong to an element of the relative cohomology group $H^*(\CM_6, \Pi_{\rm D6}, \IZ_6)$. See \cite{Marchesano:2014bia} for other applications of such relative (co)homology groups to type IIA compactifications with D6-branes.} In particular we have that
\be
d(dA_1 + \bar{G}_2)\, =\, \sum_\a  \d(\Pi_\a)  + \d(\CR\Pi_\a) - 4\d(\Pi_{\rm O6})
\ee
which has a globally well-defined solution for the two-form flux $dA_1 + \bar{G}_2$ due to the tadpole condition (\ref{RRtadpole}).  The corresponding contribution to the superpotential reads
\be
l_s W =  \frac{1}{2l_s^{5}} \int_{\CM_6} (dA_1 + \bar{G}_2) \wedge J_c \wedge J_c + \dots  = \frac{1}{2l_s^4} \int_{\Sig_4 + \Lam_4} \hspace*{-.5cm}  J_c \wedge J_c + \dots  = \oh \tilde m^a \CK_{abc}T^bT^c + \dots
\label{supotildem}
\ee
In the first equality we have used the fact that the two-form $dA_1 + \bar{G}_2$ is quantised as in (\ref{Diracq}), and that $J_c^2$ is closed to convert the integral over $\CM_6$ into an integral over the four-chain $\Sigma_4$, following \cite{Hitchin:1999fh} and also \cite{Marchesano:2014bia,Marchesano:2014iea}. As before, this chain is such that $\p \Sig_4 = \sum_\a \Pi_\a + \CR\Pi_\a - 4\Pi_{\rm O6}$, and so it is determined only up to a closed four-cycle $\Lam_4$, which can be understood as the contribution to the superpotential  coming from $W_K$.  In the second equality we have simply used the expression (\ref{tmchain}) for the dressed two-form flux $\tilde m^a$. 

In fact, this discussion provides a new interpretation of $\tilde m^a$. Indeed, let us use Hodge decomposition to split the two-form $dA_1 + \bar{G}_2$ such that $G_2$ is a purely harmonic two-from, while $dA_1$ is a sum of an exact and co-exact two-forms. Then we necessarily have that
\be
\bar{G}_2\, =\, \tilde m^a \om_a .
\ee
That is, the gauge invariant quantity $\tilde m^a$ is nothing but a harmonic component of the two-form flux, computed including the D6-brane backreaction. Notice that due to such backreaction $\tilde m^a$ does not need to be quantised, since $dA_1$ also contributes to integrals over two-cycles $\pi_2 \in \CM_6$ via its co-exact component. 

Finally, a similar analysis can be carried for the other components of the polyform $d{\bf A} + \bar{\bf G}$, obtaining the rest of the open-closed superpotential and an analogous interpretation for the remaining dressed fluxes $\tilde e_0$, $\tilde e_a$.

\section{Holomorphic variables and the K\"ahler potential}
\label{s:kahler}

As discussed in \cite{Grimm:2011dx,Kerstan:2011dy}, by dimensionally reducing the 10d type IIA supergravity and D6-brane actions one finds that the open string moduli and RR bulk axions mix kinematically. In terms of the 4d $\cn=1$ effective field theory this is interpreted as a redefinition of the chiral superfields containing such RR bulk axions, with the new holomorphic variables depending on the open string fields. This behaviour is analogous to the one observed in type IIB orientifold compactifications \cite{Antoniadis:1996vw,DeWolfe:2002nn,Angelantonj:2003zx,Grana:2003ek,Berg:2004ek,Jockers:2004yj,Baumann:2006th,Marchesano:2008rg,Camara:2009uv,BerasaluceGonzalez:2012vb,Martucci:2014ska}, and dictates how open string fields enter into the K\"ahler potential. In principle, one may determine what the new holomorphic variables are by demanding that the new K\"ahler potential reproduces the above kinetic mixing and that quantities like the gauge kinetic function depend holomorphically on the new chiral coordinates. The last requirement is however delicate to implement in generic $\cn=1$ compactifications, as it is known that loop corrections will play an important role \cite{Berg:2004ek,Berg:2005ja,Berg:2005yu,Haack:2008yb} and these are difficult to compute in general \cite{Berg:2014ama}. 

In the following we would like to apply an alternative approach, based on the discrete symmetries discussed in the previous sections. In particular we will implement in our setup the reasoning of \cite{Camara:2009uv}, in which the definition of holomorphic variables was related to their transformation under discrete shifts of the open string fields. 

To proceed we need to understand how to dualise correctly in 4d the two forms dual to the axions $\text{Re}\, N'^K$. To do so we can add to the 4d effective
action the following Lagrange multiplier
\be
\frac{1}{4\kappa_4^2}\int_{\IR^{1,3}}  d\rho_K \wedge d \,\re\, N'^K
\label{C3kin}
\ee
where $\rho_K$ are the two-forms dual to the 4d axions $\re\, N'^K$, arising from the RR five-form potential as $C_5 \, =\, \rho_K \wedge \beta^K$. Such two-forms couple to the D6-brane moduli via its Chern-Simons action and in particular through the term
\be
\mu_6 \int_{\IR^{1,3} \times {\cal C}_4^\a} \hspace*{-.25cm} dC_5 \wedge \sig \tilde{F} 
\ee
from where we obtain the four-dimensional coupling
\be
 - \frac{1}{4 \kappa_4^2} \int_{\IR^{1,3}} d\rho_K \wedge d\theta^i_\a\left( l_s^{-4} \int_{{\cal C}_4^\a} \beta^K \wedge \zeta_i\right) \, =\,  - \frac{1}{8 \kappa_4^2} \int_{\IR^{1,3}} d\rho_K \wedge  g^K_{\a\, i} d\theta^i_\a\, .
\ee
As a result, we have that the bulk axions transform as follows under discrete shifts of the open string moduli
\bea
\label{nhshiftWL}
\Phi^i_\a\, \raw\, \Phi^i_\a + k^i_\a & \qquad \qquad & \re\, N'^K \, \raw\, \re\, N'^K  \\
\Phi^i_\a\, \raw\, \Phi^i_\a - T^a \Delta f^i_{\a\, a} & \qquad \qquad &  \re\, N'^K \, \raw\, \re\, N'^K + \frac{1}{2} \theta^i_\a  \Delta g^K_{\a\, i}
\label{nhshiftpos}
\eea
where we have defined
\be
\Delta f^i_{\a\, a}\, =\, \frac{1}{l_s^{4}} \int_{\Lam_4^\a}  \om_a \wedge \tilde \rho^i \qquad \text{and} \qquad  \Delta g^K_{\a\, i}\, =\, \frac{1}{l_s^{4}} \int_{\Lam_4^\a}  \beta^K \wedge \tilde \zeta_i
\label{Deltafg}
\ee
which for a four-cycle $\Lam_4^\a$ are integer numbers. It is easy to see that these transformations are not holomorphic. However, as in \cite{Camara:2009uv} one may redefine the complex structure moduli $N'^K$ to new variables that transform holomorphically under the above shifts. Indeed, let us define
\be
N^K\, =\, N'^K - \frac{1}{2} T^a \sum_\a  {\bf H}^K_{\a\, a} 
\label{redefNK}
\ee
where ${\bf H}^K_{\a\, a}$ are real functions of the three-cycle position moduli defined by
\be
\p_{\varphi_\b^j} {\bf H}^K_{\a\, a}\, =\, (\eta_{\b\, a})^i{}_j g_{\a\, i}^K\d_{\a\b}\, .
\label{defH}
\ee
The functionals ${\bf H}^K_{\a\, a}$ are similar to those defined by Hitchin in \cite{Hitchin:1997ti} in order to describe the metric on the moduli space of special Lagrangian submanifolds,\footnote{To connect with \cite{Hitchin:1997ti} notice that (\ref{defH}) is equivalent to $\p_{\phi_\b^i} \left(t^a {\bf H}^K_{\a\, a}\right) = - g_{\a\, i}^{K} \d_{\a\b}$, with $\phi_\a^i = \im\, \Phi_\a^i$.} and were used in \cite{Grimm:2011dx} to propose a K\"ahler potential including D6-brane moduli. From the definitions of section \ref{s:typeIIA} one can see that along a periodic direction of the D6-brane position 
\be
 f^i_{\a\, a} \, =\, s\, \Delta f^i_{\a\, a} \qquad \qquad s\, \in\, \IR
\ee
the function $g^K_{\a\, i}$ must be of the form
\be
g^K_{\a\, i}\, =\, s\, \Delta g^K_{\a\, i} + M^K_{\a\, i} (s)
\ee
where we have defined  $\Delta f^i_{\a\, a}$ and $\Delta g^K_{\a\, i}$ as in (\ref{Deltafg}). Here $M^K_{\a\, i}$ is a periodic function of period one in $s$, with mean $m^K_{\a\, i}$ and such that $M^K_{\a\, i}(0) =0$. We then have that
\be
{\bf H}^K_{\a\, a}\, =\, \oh s^2\,  \Delta f^i_{\a\, a}\Delta g^K_{\a\, i} + s\,  \Delta f^i_{\a\, a} m^K_{\a\, i} + {P}^K_{\a\, a}(s)
\ee
where ${P}^K_{\a\, a}(s+1) = {P}^K_{\a\, a}(s)$. Hence along this periodic direction ${\bf H}^K_{\a\, a}$ shifts as
\be
{\bf H}^K_{\a\, a}(s+1) - {\bf H}^K_{\a\, a}(s)\, =\, f^i_{\a\, a}\Delta g^K_{\a\, i}   + \Delta f^i_{\a\, a} \left(\oh \Delta g^K_{\a\, i} +  m^K_{\a\, i}\right)
\ee
and therefore the redefined variables shift as
\bea
\label{hshiftWL}
\Phi^i_\a\, \raw\, \Phi^i_\a + k^i_\a &  \qquad & \re\, N^K \, \raw\, \re\, N^K \\ \nonumber
\Phi^i_\a\, \raw\, \Phi^i_\a - T^a \Delta f^i_{\a\, a} &  \qquad &  \re\, N^K \, \raw\, \re\, N^K +  \oh \Phi^i_\a  \Delta g^K_{\a\, i} -  \frac{1}{4}  T^a\Delta f^i_{\a\, a} \left(\Delta g^K_{\a\, i} +  2m^K_{\a\, i}\right)
\label{hshiftpos}
\eea
giving the desired holomorphic behaviour. 

The same reasoning can be applied to the complex structure moduli $U_{\Lam}'$, obtaining the redefined variables
\be
U_\Lam \, =\, U_\Lam^{\prime} + \frac{1}{2}T^a  \sum_\a {\bf H}_{\a \, \Lam\, a} \qquad \text{with} \qquad \p_{\varphi^j_\b} {\bf H}_{\a \, \Lam\, a} \, =\,  (\eta_{\b\, a})^i{}_j g_{\a \, \Lam\, i} \d_{\a\b}
\label{redefUlam}
\ee
that show the appropriate holomorphic behaviour.  

Performing this change of variables into the K\"ahler potential will implement its dependence on the open string fields. More precisely we have that the piece (\ref{KK}) remains invariant, while (\ref{KQ}) should be rewritten in terms of the new holomorphic variables. 
That is, we should again consider
\be
K_Q\, =\, - 2 \text{log} (\cg^Q(n^{\prime K},u^\prime_\Lam))
\label{KQredef}
\ee
but express $n^{\prime K} = \im\, N^{\prime K}$ and $u^\prime_\Lam = \im \,U^\prime_\Lam$  in terms of the new holomorphic variables, namely
\be
n^{\prime K} \, =\, n^{K} + \oh t^a \sum_\a {\bf H}^K_{\a\, a} \qquad \text{and} \qquad u_\Lam^{\prime} \, =\, u_\Lam -  \oh t^a \sum_\a  {\bf H}_{\a \, \Lam\, a}  
\label{nuprime}
\ee
with $n^{K} = \im\, N^{K}$, $u_\Lam = \im \,U_\Lam$ and $t^a = \im\, T^a$. Although our definition of holomorphic variables is different from the proposal in \cite{Grimm:2011dx}, it embeds the Hitchin functionals into the K\"ahler potential in a similar fashion, reproducing the same K\"ahler metrics for the open string fields. Finally, one can check that in the toroidal case this redefinition reduces to
\be
n'^K\, =\, n^K + \frac{1}{4} \sum_\a (\CQ^K_\a)_{ij} \left[(\im\, T^a\, \eta_{\a\, a})^{-1} \right]^j{}_k\, \im\, \Phi^i_{\a} \im\, \Phi^k_{\a}
\label{redefTorus}
\ee
and similarly for $u_\Lam$, in agreement with standard result in type IIB toroidal orientifolds \cite{Antoniadis:1996vw,Angelantonj:2003zx,Berg:2004ek,Camara:2009uv,BerasaluceGonzalez:2012vb}. 

Notice that our reasoning partially relies on the existence of periodic directions in the moduli space of D6-branes, and such may not exist for the position moduli in generic compactifications. In these cases the definition of the functions {\bf H} could be different, as they are not constrained by periodic position shifts. Nevertheless, as discussed in Appendix \ref{ap:details}, in order to reproduce the appropriate metrics for the chiral fields one needs a redefinition of the form (\ref{nuprime}) with the {\bf H}'s defined as above. Hence, at the level of approximation that we are working, the redefinition of the complex structure moduli seems sufficiently constrained. 

One important consequence of these results is that, due to the above redefinitions, the piece of the K\"ahler potential (\ref{KQredef}) also depends on the K\"ahler moduli, and therefore the complex structure and K\"ahler moduli spaces no longer factorise. In principle this greatly complicates the computation of quantities in the 4d effective field theory, like for instance the F-term scalar potential. One can nevertheless see that, despite this complication, the redefinition of the holomorphic variables implies several non-trivial identities for the K\"ahler metrics which will be crucial for the computations of the next section. 

For instance, recall that in (\ref{KQredef}), $\cg^Q$ is a homogeneous function of degree two on the variables $n^{\prime K} = \im\, N^{\prime K}$ and $u^\prime_\Lam = \im \,U^\prime_\Lam$. As shown in Appendix \ref{ap:details}, these two variables are in turn homogeneous functions of degree one in $\{\psi^\b\} =\{t^a, n^{K}, u_\Lam, \phi^i\}$, where for simplicity we have absorbed the D6-brane index $\a$ into the index $i$ in $\phi^i = \im\, \Phi^i_\a$. As a result, $\cg^Q$ is a homogeneous function of degree two on the variables $\{\psi^\b\}$, which in turn implies that  the full K\"ahler potential $K = K_K + K_Q$ satisfies
\be
K^{\a\bar{\b}}K_{\bar{\b}}\, =\,  -2i \im\, \Psi^\a
\label{sphomorel}
\ee
as well as
\be
K^{\a\bar{\b}}K_\a K_{\bar{\b}}\, =\,  7
\label{noscale}
\ee
where in both cases the sum is taken over all fields $\psi^\a$. 

In addition, from the same definition of the K\"ahler potential and some simplifying assumptions several important relations for the elements of the inverse K\"ahler metric follow. First we have
\be
K^{i \bar a } =  - f_{ b}^i K^{b \bar a }
\label{keyrel}
\ee
%
Second it happens that
\be
K^{a\bar{b}}\, =\, \left( \p_{a}\p_{\bar{b}} K_K \right)^{-1}
\label{invKab}
\ee
or in other words the inverse K\"ahler metric for the K\"ahler moduli is exactly the same as in the absence of open string degrees of freedom. Finally for the specific definition of {\bf H} taken above we have that
\be
K^{i \bar \jmath} -  K^{a \bar b}f_a^if_b^j\, =\, G^{ij}_{\rm D6}
\label{invKij}
\ee
where $G^{ij}_{\rm D6}$ is the inverse of the natural metric for one-forms in the three-cycle $\Pi_\a$
\be
G_{ij}^{\rm D6} \, = \, \frac{e^{-\phi/4}}{8 \hat{V}_6} l_s^{-3} \int_{\Pi_\a} \zeta_i \wedge *\, \zeta_j
\label{gD6}
\ee
In fact, we have that (\ref{invKij}) fixes {\bf H} up to a linear function on the $\phi^i$'s, a freedom that can be used to redefine the reference cycles $\Pi_\a^0$. We refer the reader to Appendix \ref{ap:details} for further details on all these identities.

\section{The scalar potential from 4d supergravity}
\label{s:spot}

In this section we combine all the results from the previous sections together. In particular we will show that if we take the superpotential of section \ref{s:supo} and the K\"ahler potential of section \ref{s:kahler}, we derive the scalar potential of section \ref{s:3forms} from the usual F-term 4d supergravity expression
\be
V_{\rm F}\, =\, \frac{e^K}{\kappa_4^2} \left( K^{\a \bar \b} D_\a {W} D_{\bar \b} \bar{W}- 3|W|^2\right)
\ee
where $\a$ runs over all the fields $\{\Psi^\a\} =\{T^a, N^{K}, U_\Lam, \Phi^i\}$ the K\"ahler potential depends on. Now using the relations (\ref{sphomorel}) and (\ref{noscale}) we can rewrite things as
\be
K^{\alpha \bar \beta} D_\alpha W D_{\bar \beta} \overline W - 3 |W|^2 \, = \, K^{\alpha \bar \beta} \p_\alpha W \p_{\bar \beta} \overline W+ 4 \text{Im}  \left(
\text{Im}\, \Psi^\alpha \p_\alpha W \overline W\right) + 4 |W|^2\,,
\label{splitVF}
\ee
and then analyse this expression term by term. 

First we have that
\be\begin{split}
4 l_s^2 |W|^2 &= \bigg[2 e'_a t^a + 2 \mathcal K_{ab} m'^a b^b - \frac{1}{3} m \mathcal K + m \mathcal K_{ab} b^a b^b -
2 \text{Re} \, \Phi^i n_{a\, i} t^a + 2 \text{Im} \, \Phi^i (n_{Fi} - n_{a\, i} b^a)\bigg]^2  
\\& + \bigg[2 e'_0 + 2 e_a' b^a + \mathcal K_{abc} m'^a b^b b^c - \mathcal K_a m'^a + \frac{1}{3} m \mathcal K_{abc} b^a b^b b^c
- m \mathcal K_{a} b^a + 2 \text{Im} \,\Phi^i n_{a\, i}t^a \\ & + 2 \text{Re} \, \Phi^i (n_{F\,i} - n_{a\, i} b^a) \bigg]^2
\end{split}\ee
where we have defined
\be
\CK\, =\, \CK_{abc} t^at^bt^c \qquad \CK_a\, =\, \CK_{abc} t^bt^c \qquad \CK_{ab}\, =\, \CK_{abc} t^c
\label{Ks}
\ee
and  have merged the contribution of $W_{\rm D6}^0$ in (\ref{supoD6full}) and the closed string piece of the superpotential $W_K$ (\ref{WK}) by defining the primed fluxes
\be
e'_0\, =\, e_0 + \frac{1}{8\pi^2} \int_{\Sig_4^0} \tilde F \wedge \tilde F \qquad  e'_a\, =\, e_a - l_s^{-2} \int_{\Sig_4^0} \om_a \wedge \frac{\tilde F}{2\pi} \qquad m'^a \, =\, m^a + l_s^{-4} \int_{\Sig_4^0} \tilde\om^a
\ee
Second we have that
\be\begin{split}
4l_s^2 \text{Im} \left[\psi^\alpha \p_\alpha W \overline W\right] =& 
-4 \text{Im}\, W\left[t^a(e'_a + \mathcal K_{abc}m'^b b^c + \frac{1}{2}m \mathcal K_{abc}b^b b^c - \frac{1}{2}m \mathcal K_{a}-\text{Re}\, \Phi^i n_{a\, i})\right. \\ 
& +\left.\text{Im} \, \Phi^i (n_{Fi} - n_{a\, i}b^a)\right] + 4 \text{Re}\, W \left[t^a(\mathcal K_{ab} m'^b + m \mathcal K_{ab} b^b -2 \text{Im} \,\Phi^i n_{a\, i})\right]
\end{split}
\ee
Hence, summing the two results we find 
\be
\label{eq:1}
\begin{split}
4 l_s^2 |W|^2+4 l_s^2\text{Im} \left[\psi^\alpha \p_\alpha W \overline W\right] 
=&\bigg[2 e'_0 + 2 e'_a b^a + \mathcal K_{abc} m'^a b^b b^c  + \frac{1}{3} m \mathcal K_{abc} b^a b^b b^c
+2 \text{Re} \, \Phi^i (n_{F\, i} - n_{a\, i} b^a) \bigg]^2\\
& - \bigg[\mathcal K_a m'^a + m \mathcal K_a b^a - 2 \text{Im} \, \Phi^i n_{a\, i} t^a\bigg]^2  + \frac{4}{3} m \,\mathcal K \,\text{Im}\, W
\end{split}
\ee

Finally, from the relation (\ref{keyrel}) we can express the remaining term in (\ref{splitVF}) as
\be
K^{\alpha \bar \beta} \p_\alpha W \p_{\bar \beta} \overline W = K^{a \bar b} \left[\p_a W - f_a^i \p_i W\right]
\left[\p_{\bar b} \overline W -f_b^i \p_{\bar \imath}\overline  W\right] + (K^{i \bar \jmath}
-  K^{a \bar b}f_a^i f_b^j) \p_i W \p_{\bar \jmath} \overline W
\ee
where here $a$, $b$ run over the K\"ahler moduli and $i$, $j$ over the open string moduli that appear in the superpotential. This implies that 
\be\label{eq:2}\begin{split}
l_s^2 K^{\alpha \bar \beta} \p_\alpha W \p_{\bar \beta} \overline W=& \,
K^{a \bar b} \bigg[e'_a + \mathcal K_{acd } m'^c b^d + \frac{1}{2} m \mathcal K_{acd} b^c b^d -\frac{1}{2} m\mathcal K_{a} - \text{Re} \,\Phi^i n_{a\, i} - f_a^i (n_{F\, i} - n_{c\, i} b^c)\bigg]\\
&\quad \times \bigg[e'_b + \mathcal K_{bcd } m'^c b^d + \frac{1}{2} m \mathcal K_{bcd} b^c b^d -\frac{1}{2}m \mathcal K_{b} - \text{Re} \,\Phi^i n_{b\, i} - f_b^i (n_{F\, i} - n_{c\, i} b^c)\bigg]\\
&+K^{a \bar b} \bigg[\mathcal K_{ac} m'^c + m \mathcal K_{ac} b^c - \text{Im} \,\Phi^i n_{a\, i} + f_a^i  n_{c\, i} t^c \bigg]\\
&\quad \times \bigg[\mathcal K_{bc} m'^c + m \mathcal K_{bc} b^c - \text{Im} \,\Phi^i n_{b\, i} + f_b^i  n_{c\, i} t^c \bigg]\\
&+\big[K^{i \bar \jmath}- K^{a \bar b}f_a^if_b^j\big] \p_i W \p_{\bar \jmath} \overline W\,.
\end{split}\ee

Using (\ref{pullbackid}) we can simplify the first term in the second line of \eqref{eq:1} to
\be
-\Big[\mathcal K_a (m'^a + m b^a + q^a)\Big]^2\, =\, - \Big[\mathcal K_a (\tilde m^a + m b^a)\Big]^2
\ee
as well as the terms appearing in the third and fourth lines of \eqref{eq:2}, that now read
\be
K^{a \bar b} \mathcal K_{a d} \mathcal K_{b c} \Big[\tilde m^d + m b^d\Big]\Big[\tilde m^c + m b^c\Big] 
\ee
where we have expressed everything in terms of the dressed fluxes $\tilde m^a = m'^a + q^a$. Adding these last two terms we obtain
\be
\left( K^{a \bar b} \mathcal K_{a d} \mathcal K_{b c} +\CK_d\CK_c\right) \Big[\tilde m^d + m b^d\Big]\Big[\tilde m^c + m b^c\Big] = 16 e^{3\phi} \hat{V}_6^2 g_{dc} \Big[\tilde m^d + m b^d\Big]\Big[\tilde m^c + m b^c\Big]
\ee
where we have used (\ref{invKab}) and more precisely (\ref{ap:invKab}).

The remaining terms can be arranged as follows. One may first rewrite the first two lines of the rhs of \eqref{eq:2} as
\be
\label{eq:3}
\begin{split}
K^{a \bar b}& \bigg[\tilde e_a + \mathcal K_{acd } \tilde m^c b^d + \frac{1}{2} m \mathcal K_{acd} b^c b^d\bigg] \times \bigg[\tilde e_b + \mathcal K_{bcd } \tilde m^c b^d + \frac{1}{2} m \mathcal K_{bcd} b^c b^d  \bigg]\\
-\,&K^{a \bar b}m\mathcal K_{a}\bigg[\tilde e_b + \mathcal K_{bcd} \tilde m^c b^d + \frac{1}{2} m \mathcal K_{bcd} b^c b^d \bigg]
+\frac{1}{4} K^{a \bar b} \mathcal K_a \mathcal K_b m^2
\end{split}
\ee
where we have again used (\ref{pullbackid}), the definition of $\tilde m^a$ and that $\tilde e^a = e'^a  - n_{a\, i} \th^i - n_{F\, i} f^i_a$. Then, by using that $K^{a \bar b} \mathcal K_a = \frac{4}{3}\CK t^b$ we can add the second line of this equation to the last term in \eqref{eq:1} and obtain
\be
\begin{split}
& \frac{4}{3}\CK m \left[\left( \tilde e_at^a + \CK_{ab}\tilde m^ab^b -\frac{1}{6} m\CK + \oh m \CK_{ab}b^ab^b \right) - t^b \left( \tilde e_b + \CK_{bcd}\tilde m^cb^d + \oh \CK_{bcd}b^cb^d\right)\right]  \\ 
&+\frac{1}{4} K^{a \bar b} \mathcal K_a \mathcal K_b m^2\, =\, \left(\frac{1}{3} - \frac{2}{9}\right) \CK^2 m^2\, =\, 4 e^{3\phi} \hat{V}_6^2  m^2
\end{split}
\ee
Summing all these contributions we find the following F-term scalar potential 
\bea\nonumber
 V_F \, = \, \frac{e^K}{l_s^2 \kappa_4^2} &&\Bigg\{\left[ 2 \tilde e_0 + 2 \tilde e_a b^a + \mathcal K_{abc} \tilde m^a b^b b^c  + \frac{1}{3} m \mathcal K_{abc} b^a b^b b^c\right]^2 \\ \nonumber
 &  & +\,  g^{ab} \left( \tilde e_a +\CK_{acd}\tilde m^cb^d + \frac{m}{2} \CK_{acd}b^cb^d\right) \left(\tilde e_b +\CK_{bef} \tilde m^eb^f + \frac{m}{2} \CK_{bef}b^eb^f\right) \\
 &  & +\, \frac{4}{9} \CK^2 g_{ab} \left( \tilde m^a + mb^a \right)  \left( \tilde m^b + mb^b \right)  \, +\, \frac{1}{9} \CK^2 m^2\Bigg\} +V_{\rm DBI}
\eea
where we have defined
\be
V_{\rm DBI}\,  =\,  \frac{e^K}{\kappa_4^2} \big[K^{i \bar \jmath}
-  K^{a \bar b}f_a^if_b^j\big] \p_i W \p_{\bar \jmath} \overline W\,.
\label{VDBI}
\ee
Hence, we do indeed recover an F-term scalar potential which is the sum of two terms. The first one is the potential $V_{{\rm RR} + {\rm CS}}$ computed in section \ref{s:3forms} which has the form of the usual type IIA scalar potential generated by RR fluxes but with those replaced with the dressed fluxes $\{\tilde e_0, \tilde e_a, \tilde m^a, m\}$ that contain the open string moduli dependence. The second piece (\ref{VDBI})  should then correspond to the DBI contribution to the F-term scalar potential, which is non-trivial when $\sig F - J_c$ does not vanish over $\Pi_\a$. Whenever such source of supersymmetry breaking is small in string units, the corresponding excess of energy is given by \cite{Escobar:2015ckf}
\bea
V_{\rm DBI}  & = & \frac{e^K}{l_s^2 \kappa_4^2} 8 \hat{V}_6 e^{\phi/4} \frac{1}{l_s^{3}}\int_{\Pi_\a} \left( \sig F - J_c\right)\wedge * \left( \sig F - \overline{J_c}\right) \\
&= & \frac{e^K}{l_s^2 \kappa_4^2} G^{ij}_{\rm D6} \left(n_{F\, i} - n_{a\, i} T^a\right)\left(n_{F\, j} - n_{a\, j} \bar{T}^a\right)   
\eea
where $G^{ij}_{\rm D6}$ is the inverse of (\ref{gD6}). We then see that the choice of K\"ahler metric taken in the previous section also reproduces the expected contribution from the DBI action to the F-term potential.

\section{Type IIB models with D7-brane Wilson lines}
\label{s:typeIIb}

In this section we translate our results obtained in type IIA compactifications with O6-planes to type IIB compactifications with O7/O3-planes.
We consider compactifications of type IIB string theory on $\mathbb R^{1,3} \times \mathcal M_6$ where $\mathcal M_6$ is taken to be a 
compact Calabi-Yau 3-fold. The orientifold action is given by the $\Omega_p (-1)^{F_L} \mathcal R$ where in this case 
$\mathcal R$ is a holomorphic involution of the Calabi-Yau manifold. In particular the action of $\mathcal R$ on the K\"ahler form $J$ and the 
holomorphic 3-form $\Omega$ of the Calabi-Yau 3-fold $\mathcal M_6$ is 
\be
\mathcal R J = J \,, \quad \mathcal R \Omega = - \Omega\,.
\ee
At the fixed loci of this involution which can be either points or 4-cycles we find O3-planes and O7-planes respectively. Having introduced an orientifold 
involution it is necessary to cancel the total RR charge induced by the orientifold planes in the compact space. To cancel the RR tadpole induced by the 
O7-planes we may introduce spacetime filling D7-branes wrapping 4-cycles $\mathcal S_a$ in $\mathcal M_6$ satisfying the homological relation
\be
\sum_a [\mathcal S_a]+ [\mathcal R \mathcal S_a ] = 8\,[\pi_{O7}]\,.
\ee
In a similar fashion the tadpole induced by the O3-planes may be cancelled by introducing an adequate number of D3-branes. 

Preservation of 4d $\mathcal N=1$ supersymmetry forces us to take the cycles $\mathcal S_a$ to be holomorphic divisors of $\mathcal M_6$ with the 
worldvolume flux $\mathcal F = B|_{\mathcal S_a} - \sigma F$ satisfying the conditions 
\be
\mathcal F^{(0,2)} = 0 \,, \quad J \wedge \mathcal F =0\,,
\ee
where $\alpha^{(p,q)}$ stands for the Hodge type of the differential form $\alpha$. 

The 4d effective action for the closed string sector contains the axio-dilaton $\tau = C_0 + i e^{-\phi}$, K\"ahler moduli and complex structure moduli, with a factorised moduli space, see \cite{Grimm:2004uq}.
The K\"ahler moduli are obtained from the dimensional reduction of the K\"ahler form $J$, of the B-field and of the RR forms $C_2$ and $C_4$ as
\bea
J &= v^\alpha\, \omega_\alpha \,,\qquad &\omega_\alpha \in H^{2}_+ (\mathcal M_6,\mathbb Z)\,,\nonumber\\
B &= b^a\, \omega_a \,,\qquad&\nonumber\\
C_2 &= c^a\, \omega_a \,,\qquad &\omega_a \in H^{2}_- (\mathcal M_6,\mathbb Z)\,,\nonumber\\
C_4 &= C_\alpha\, \tilde \omega^\alpha\,, \qquad& \tilde \omega^\alpha \in H^{4}_+ (\mathcal M_6,\mathbb Z)\,.
\eea
With these definitions at hand we may define the following 4d chiral multiplets
\be
G^a = c^a - \tau b^a\,,\quad 
T_\alpha = \frac{1}{2} \mathcal K_{\alpha \beta \gamma} v^\beta v^\gamma - \frac{i}{2(\tau - \overline \tau)}\mathcal K_{\alpha b c } G^b (G - \overline G)^c - i C_\alpha\,,
\ee
where we defined the triple intersection numbers
\be
\mathcal K_{\alpha \beta \gamma} = \frac{1}{l^6_s} \int_{\mathcal M_6} \omega_\alpha \wedge \omega_\beta \wedge \omega_\gamma\,, 
\quad \mathcal K_{\alpha b c} = \frac{1}{l^6_s} \int_{\mathcal M_6} \omega_\alpha \wedge \omega_b \wedge \omega_c\,.
\ee
The K\"ahler potential at large volume for these moduli is then written as an implicit function of the chiral multiplets as 
\be\label{eq:KK}
K_K = - 2 \log \left[\frac{1}{6} \mathcal K_{\alpha \beta \gamma} v^\alpha v^\beta v^\gamma\right]\,.
\ee
For an explicit expression it is necessary to invert the relation between the volume of 2-cycles $v^\alpha$ 
and the chiral coordinates $T_\alpha$. While this is hard to do in general, one can see that $e^{-K_K}$ is a homogeneous function of degree three on the moduli. 

The complex structure moduli are obtained by performing the dimensional reduction of the holomorphic 3-form $\Omega$ on harmonic 3-forms. 
Due to the orientifold projection we only need to consider 3-forms odd under the orientifold involution. After taking $(\alpha_I, \beta^I)
$ as a symplectic basis for $ H^3_-(\mathcal M_6,\mathbb Z)$ we obtain
the following expansion for 
$\Omega$
\be
\Omega = X^I \,\alpha_I - F_I \,\beta^I \,.
\ee
The quantities $(X^I, F_I)$ are called periods of the holomorphic 3-form $\Omega $ and depend on the complex structure of $\mathcal M_6$. In particular
it  is possible to prove that the $F_I$ are functions of the $X^I$ implying that the latter may be chosen as good projective coordinates on the complex
structure moduli space. Fixing $X^0 = 1$ we will identify the remaining $X^I $ with the complex structure moduli with $I = 1,\dots , h^{(2,1)}_- (\mathcal
M_6)$. Additionally it is possible to prove that the $F_I = \p_I F $ where the function $F$, usually called the prepotential, 
is homogeneous  of degree 2 in the 
projective coordinates $X^I$. The K\"ahler potential for the complex structure moduli may be written using the periods
\be
K_Q = - \log\left[i (X^I \overline F_I - \overline X^I F_I)\right] \,.
\ee
Finally we need to consider the axio-dilaton whose K\"ahler potential is simply
\be\label{eq:Ktau}
K_{\tau} = -\log\big[-i(\tau - \overline \tau)\big]\,.
\ee
The introduction of an open string sector will add moduli in the 4d effective field theory. The case of D7-branes was analysed in \cite{Jockers:2004yj} where it was found that for 
D7-branes the moduli fall in two separate classes, brane position moduli and Wilson line moduli. In \cite{Jockers:2004yj} it was found that brane position moduli, counted by the 
cohomology group $H^{(2,0)}(\mathcal S_a)$, give a redefinition of the axio-dilaton and therefore modify the K\"ahler potential \eqref{eq:Ktau}. Similarly it was found 
that Wilson line moduli, counted by the cohomology group $H^{(0,1)}(\mathcal S_a)$, enter in a redefinition of the K\"ahler moduli thus modifying the K\"ahler potential 
\eqref{eq:KK}. In the following we shall revisit the definition of the K\"ahler potential for the Wilson line moduli.

\subsubsection*{Adding Wilson lines}

The definition of Wilson line moduli for a D7-brane on a 4-cycle $\mathcal S_a$ is relatively straightforward, as they come from the dimensional reduction of the D7-brane gauge 
field on elements of $H^1(\mathcal S_a)$. To correctly write 4d chiral fields it is necessary to perform the dimensional reduction using elements of $H^{(0,1)}(\mathcal S_a)$
giving the following definition for the Wilson line moduli
\be
\xi^{a}_i = \frac{2}{l_s^5}\int_{\mathcal S_a} \sig A \wedge \chi_i \,, 
\ee
where $\chi_i $ are elements of $H^{(2,1)}(\mathcal S_a)$. According to this definition the internal profile for a Wilson line scalar will be proportional to a suitable element 
of $H^{(1,0)}(\mathcal S_a)$ leading to the following dimensional reduction ansatz for the gauge field on the D7-brane $a$
\be\label{eq:WL1}
A =\frac{\pi}{l_s} \text{Im} [\xi^{a}_i \, \overline \gamma^i ]\,,
\ee
where the differential forms $\overline \gamma^i$ give a basis of $ H^{(0,1)}(\mathcal S_a)$. At this point it is important to choose properly the normalisation of the differential
forms $\gamma^i$ to ensure that the axionic components of the Wilson line moduli have a fundamental period independent on the complex structure on $\mathcal S_a$.
To this end it is convenient
to introduce a suitable basis  $(\tilde \alpha_i , \tilde \beta^j)$ of $H^1(\mathcal S_a,\mathbb Z)$ which allows us to express the differential forms $ \gamma^i$
as\footnote{We note that when considering F-theory compactifications a similar ansatz was made in \cite{Grimm:2010ks,Grimm:2015ona,Greiner:2015mdm} for the 3-forms
of the Calabi-Yau fourfold. This comes as no surprise as the Wilson line moduli of D7-branes come as 3-forms when considering the uplift to F-theory.}
\be
 \gamma^{i} = (\text{Im}\, f^{a})^{ij}\left[\tilde \alpha_j +  f^{a}_{jk}(z) \,\tilde \beta^k \right]\,,
\ee
with $f^{a}_{ij}(z)$ holomorphic in the complex structure moduli and $(\text{Im}\, f^{a})^{ij}$ is the inverse of its imaginary part. Setting $\xi^{a}_i = \eta_i +  f^{a}_{ij}(z) \theta{}^j$ we obtain that the dimensional reduction ansatz for the gauge field 
\eqref{eq:WL1} becomes
\be\label{eq:gaugeKK}
A =\frac{\pi}{l_s} \text{Im} [\xi^{a}_i \, \overline \gamma^i ]= \frac{\pi}{l_s} \left[- \eta_i\, \tilde \beta^i + \theta{}^i \tilde \alpha_i\right]\,,
\ee
implying that the fundamental periods for the axionic fields $(\eta_i,\theta^i)$ is $ \eta_i \sim  \eta_i+ 1$ and $ \theta^i \sim  \theta^i+ 1$, independent on the complex structure
moduli. As already anticipated the presence of Wilson line moduli will give a redefinition of the K\"ahler moduli of the Calabi-Yau threefold. The resulting redefinition is the following
\be\label{eq:That}
\hat T_\alpha = T_\alpha -\frac{i}{4} \sum_a (\mathcal C^{a}_{\alpha})^i_k (\text{Im}\, f^{a})^{kj} \,\xi^{a}_i \,\text{Im}\, \xi^{a}_j\,,
\ee
where we introduced the matrix $(\mathcal C^{a}_{\alpha})^i_j = l_s^{-4} \int_{\mathcal S_a} \omega_\alpha \wedge\tilde  \alpha_j \wedge \tilde \beta^i$.
To motivate this redefinition we can follow a similar logic as the one followed in section \ref{s:kahler}. 
Following \cite{Jockers:2004yj} we add the
following Lagrange multiplier term in the effective action 
\be
-\frac{1}{4 \kappa_4^2} \int_{\mathbb R^{1,3} }d \rho_2^\alpha \wedge d C_\alpha \,,
\ee
where $\rho^\alpha $ are duals of the RR axions $C_\alpha$ in 4d coming from the dimensional reduction of $C_4 $  along 2-forms $\omega_\alpha$.
These two forms have additional couplings to the Wilson line moduli of the D7-branes coming from the term
\be
\frac{\mu_7 }{2}\int_{\mathbb R^{1,3} \times \mathcal S_a} C_4 \wedge \mathcal F \wedge \mathcal F\,.
\ee
in the Chern-Simons action of the D7-brane. Performing dimensional reduction of this term we obtain
\be
-\frac{1}{16\kappa_4^2} \int_{\mathbb R^{1,3}} d\rho^\alpha\wedge  \text{Im} \left[\bar \xi^{a}_i \,d \xi^{a}_j\right] (\text{Im} f^{a})^{ik} (\mathcal C^{a}_\alpha)^{j}_k\,,
\ee
which closely resembles the result obtained in \cite{Camara:2009uv}. Following the discussion there, it is possible to see 
that the variable \eqref{eq:That} shifts holomorphically under discrete shifts of the Wilson line moduli.

\subsubsection*{Superpotential and discrete symmetries}

When turning on RR fluxes the superpotential obtained in the 4d effective field theory has the simple form
\be
l_s W_{\rm F} = X^I e_I +F_I\, m^I\,,
\ee
where the closed string flux $F_3$ has been expanded as
\be
l_s^{-2} F_3 = m^I \alpha_I + e_I \,\beta^I \,.
\ee
Introduction of D7-branes will add some terms in the superpotential. Specifically the D7-brane superpotential is \cite{Jockers:2004yj,Martucci:2006ij}
\be
l_s W_{\rm D7} =\frac{1}{ l_s^{5}}\int_{ \Sigma_5} \Omega \wedge \mathcal F
\ee
where, similarly to the case considered in type IIA, $\Sigma_5$ is the five-chain connecting all D7-branes to the O7-planes, and $\CF$ an extension of the worldvolume flux in it. Splitting the five-chain as
$\Sigma_5 = \Sigma_5^0 + \sum_{a} \mathcal C_5^{a}+\mathcal  R \mathcal C_5^{a}$ we obtain the following D7-brane superpotential
\be
l_s W_{\rm D7} =\sum_{a}\frac{2}{ l_s^{5}}\int_{ \mathcal C_5^a} \Omega \wedge \mathcal F+\frac{1}{l_s^{5}}\int_{ \Sigma^0_5} \Omega \wedge \mathcal F=\sum_{a}\frac{2}{ l_s^{5}}\int_{ \mathcal C_5^a} \Omega \wedge \mathcal F+l_s W_{\rm D7}^0\,.
\ee
where the index $a$ runs over the D7-branes but not their orientifold images.
In the following we shall focus our attention on the dependence on Wilson line moduli thus neglecting the presence of brane position moduli in the 
superpotential. In this case the superpotential becomes 
\be\label{eq:supoD7}
l_s W_{\rm D7} = \sum_a \frac{2}{l_s^5} \int_{\mathcal C_5^a} \Omega \wedge \sigma \tilde F +l_s W_{\rm D7}^0 = -\sum_a \frac{2}{l_s^4} \left[\int_{\mathcal S_a}  \Omega \wedge \sigma A -\int_{\mathcal S^0_a}  \Omega \wedge \sigma A \right]
+ l_s W_{\rm D7}^0\,.
\ee
Choosing the configuration of the D7-branes such that on the reference 4-cycles $\mathcal S_a^0$ Wilson lines are turned off it is possible to drop 
the last two  terms in \eqref{eq:supoD7}. At this point we may use \eqref{eq:gaugeKK} to write the superpotential in terms of the Wilson line scalars
as 
\be
l_s W_{\rm D7} = \sum_a \eta_i \left[(c_{a})^i_I X^I-(h_{a})^{i I} F_I\right]  + \theta{}^i \left[(d_{a})_i^I\, F_I - (p_{a})_{i I}\, X^I \right]\,,
\ee
where we have defined the following integer numbers
\bea
(c_{a})_I^i &=\, l_s^{-4} \int_{\mathcal S_a} \alpha_I \wedge \tilde \beta^i\,, \quad  (d_{a})_{i}^I &=\, l_s^{-4} \int_{\mathcal S_a} \beta^I \wedge \tilde \alpha_i\,,\nonumber\\
(h_{a})^{iI} &=\, l_s^{-4} \int_{\mathcal S_a} \beta^I\wedge \tilde \beta^i\,, \quad  (p_{a})_{i I} &=\, l_s^{-4} \int_{\mathcal S_a} \alpha_I \wedge \tilde \alpha_i\,.
\eea
By inspection of the D7-brane superpotential we observe that the combined flux and brane superpotential may be nicely written in terms of dressed
fluxes
\be
l_s W_{\text{IIB}} = l_s W_{\rm F} + l_s W_{\rm D7} = X^I \tilde e_I +F_I\, \tilde m^I\,,
\ee
where the redefined flux quanta are
\bea
\tilde e_I &=& e_I + \sum_a (c_{a})^{i}_I \, \eta_i - (p_{a})_{iI} \theta^i\,,\nonumber\\
\tilde m^I &=& m^I + \sum_a (d_{a})_{i}^I \, \theta^i - (h_{a})^{i I} \eta_i \,.
\label{tildedfluxIIB}
\eea
This  demonstrates that the superpotential is invariant under the discrete shifts
\bea
\eta_i \, &\raw& \,  \eta_i+ k_i\,, \qquad \left\{\begin{array}{l}e_I \, \raw \, e_I - (c_{a})^{i}_I \,k_i\\ m^I \, \raw \,
 m^I + (h_{a})^{iI} \,k_i\end{array}\right. \,, \nonumber\\
\theta^i \, &\raw& \, \theta^i+ k^i\,, \qquad \left\{\begin{array}{l}e_I \, \raw \, e_I + (p_{a})_{iI} \,k^i\\ m^I \, \raw \,
 m^I - (d_{a})_i^I \,k^i\end{array}\right.\,,
\eea
in a quite analogous fashion to their type IIA counterparts (\ref{shiftWL}) and (\ref{shiftpos}).

While this description has the advantage of making manifest the discrete symmetries of the superpotential, it is not obvious that the 
superpotential is a holomorphic function of the Wilson line moduli. To write the brane superpotential as a holomorphic function of the Wilson line moduli it is necessary to impose the following condition\footnote{Note that the condition \eqref{eq:fdef} is necessary to have $\int_{\mathcal S_a} \Omega \wedge \gamma^i = 0$ thus ensuring that the differential forms $\gamma^i $ are of Hodge type (1,0) even when $\Omega|_{\cs_4}$ is non-trivial.}
\be\label{eq:fdef}
\left[X^I (c_{a})^{i}_I- F_I \,(h_{a})^{iI}\right]   \,f^{a}_{ij} (z) = \left[F_I\, (d_{a})_{j}^I-X^I\, (p_{a})_{jI}\right]
\ee
which fixes the function $f^a_{ij} (z)$ for those Wilson lines that appear in the superpotential.

\subsubsection*{Scalar potential}

Knowing the form of the superpotential and the holomorphic variables we may perform a similar computation to the one in section section \ref{s:spot} to derive the F-term scalar potential.
For simplicity we will work in the large complex structure limit where the prepotential takes the form
\be
F = \frac{\kappa_{IJK}}{3!}  \,\frac{X^I X^J X^K}{X^0}\,,
\ee
where from now on $I=1,\dots,h^{(2,1)}_- (\mathcal M_6)$
and moreover we will write the complex structure moduli in terms of their real and imaginary parts as $z^I = X^I/X^0 = u^I + i w^I$. In addition, we will absorb the D7-brane index $a$ into the Wilson line index $i$. 
We start by noting that in this limit the relations \eqref{sphomorel},  \eqref{noscale} continue to hold if we assume that $f_{ij}^a$ is a linear function on the complex structure moduli, which we shall assume henceforth.\footnote{As shown in Appendix \ref{ap:details} these properties rely solely on $e^{-K}$ being a homogeneous function in either the real or imaginary part of the moduli, which is true whenever $\im\, f_{ij}^a$ is a homogeneous function of degree one on the $w$'s. This can be easily achieved by assuming that $f_{ij}^a$ is linear, and in the particular case of large complex structure by setting $p_a = h_a =0$ and $(c_a)^i_0 = (d_A)_i^0 = 0$, which we assume below. It would be interesting to explore more general cases. This seems compatible with the 
results of \cite{Greiner:2015mdm} where it was found that the function $f_{ij}^a$ is linear in the large complex structure limit.} The relation \eqref{keyrel} is then replaced by
\be
K^{i \bar J } = \frac{\p \text{Im}\, \xi_i}{\p \text{Im}\,z^K} \,K^{K \bar J}\,,
\ee
Using these properties it is possible to see that 
\be
K^{A\bar B} D_A W D_{\bar B} \overline W - 3 |W|^2 = K^{A\bar B} \p_A W \p_{\bar B} \overline W + 4 \text{Im}\, (\text{Im}\, \Psi^A \p_A 
W \overline W) + 4 |W|^2 \,.
\ee 
where the index $A$ runs over all fields $\Psi^A$. 
Since the remaining of the computation is very similar to the one considered in section \ref{s:spot}
we will omit most details. We find that
\bea
4l_s^2  |W|^2 +4 l_s^2\, \text{Im}\left[ \psi^A \p_A W \overline W\right] &=&  4\Big[e_0 +e_I u^I +\frac{1}{2}m^I\kappa_{IJK} u^J u^K-\frac{1}{6}m^0 \kappa_{IJK} u^I u^J u^K- \text{Re}\, \xi_i \, c^i_I u^I\Big]^2
\nonumber\\&-&\frac{4}{3} m^0 \kappa \,\text{Im}\, W-4\Big[
c_I^i w^I\, \text{Im}\, \xi_i - \frac{1}{2}m^I \kappa_{I}+\frac{1}{2} m^0 \kappa_{I} u^I \Big]^2
\eea
where analogously to the type IIA case we have defined $\kappa_{IJ} = \kappa_{IJK} w^K$, $\kappa_{I} = \kappa_{IJK} w^J w^K$, $\kappa =\kappa_{IJK} w^I w^J w^K$. Moreover we obtain that
\be
K^{A \bar B}\p_A W \bar \p_{\bar B} \overline W = \mathfrak R_I \,K^{I \bar J}\, \mathfrak R_{\bar J}+\mathfrak I_I \,K^{I \bar J}\, \mathfrak I_{\bar J}+(K^{i \bar \jmath }- 	K^{I \bar J} \p_I \, \text{Im} \, \xi_i\, \bar \p_{\bar J} \, \text{Im} \, \xi_j)\, \p_i W \bar \p_{\bar \jmath}
\overline W
\ee
where
\be
\mathfrak R_I = e_I + m^J \kappa_{IJK} u^K -\frac{1}{2} m^0 \kappa_{IJK} u^J u^K +\frac{1}{2} m^0\kappa_I -\text{Re}\, \xi_i\, c^i_I
- \p_I \text{Im} \, \xi_i\, c^i_J u^J
\ee
\be
\mathfrak I_I =  m^J \kappa_{IJ} -\frac{1}{2} m^0 \kappa_{IJ} u^J -\p_I \text{Im} \, \xi_i\, c^i_J\, w^J\,.
\ee
Summing up all contributions we obtain the final form for the scalar potential
\bea
\label{VFIIB}
 V_F \, = \, \frac{e^K}{l_s^2 \kappa_4^2} &&\Bigg\{\left[ 2  e_0 + 2 \tilde e_I u^I + \kappa_{IJK} \tilde m^I u^J u^K  -\frac{1}{3}  m^0 \kappa_{IJK} u^I u^J u^K\right]^2 \\ \nonumber
 &  & +\,  g^{IJ} \left( \tilde e_I +\kappa_{IKL}\tilde m^K u^L -\frac{1}{2}m^0 \kappa_{IKL}u^Ku^L\right) \left(\tilde e_J +\kappa_{JMN} \tilde m^Mu^N -\frac{1}{2}m^0 \kappa_{JMN}u^Mu^N\right) \\ \nonumber
 &  & +\, \frac{4}{9} \kappa^2  g_{IJ} \left(\tilde  m^I -m^0u^I \right)  \left( \tilde m^J -m^0u^J \right)  \, +\, \frac{1}{9} \kappa^2 (m^0)^2\Bigg\} +V_{\rm DBI}
\eea
where the modified flux quanta are given by (\ref{tildedfluxIIB}) and we defined 
\be
V_{\rm DBI} = \frac{e^K}{\kappa_4^2}(K^{i \bar \jmath }-K^{a \bar b} \p_a \, \text{Im} \, \xi_i\, \bar \p_{\bar b} \, \text{Im} \, \xi_j)\, \p_i W \bar \p_{\bar \jmath}
\overline W\,.
\ee
Similarly to the type IIA case, one could have arrived to the piece within brackets in (\ref{VFIIB}) by a very simple procedure. First  computing the scalar potential for the closed string modes as if there were no D7-branes, and second substituting the RR flux quanta by the quantities (\ref{tildedfluxIIB}) including the Wilson lines. Notice in fact that, because such a combination of fluxes and Wilson lines is fixed by a discrete gauge symmetry of the compactification, the same prescription applies for a general type IIB flux compactification with O3/O7-planes. Indeed, following \cite{Giddings:2001yu} one may compute the scalar potential for compactifications with RR and NS three-form fluxes and at arbitrary regions of complex structure, and then simply substitute the RR flux quanta by (\ref{tildedfluxIIB}) to obtain the scalar potential including complex structure moduli and Wilson lines. It would be interesting to elucidate how this observation constrains the K\"ahler potential for closed and open string modes, an endeavour which we plan to undertake in the future.

\section{Conclusions}
\label{s:conclu}

In this paper we have made a general analysis of the scalar potential that simultaneously involves open and closed string modes in type II Calabi-Yau compactifications with fluxes and D-branes. We have mostly focused in type IIA flux compactifications with D6-branes, and analysed the scalar potential generated at tree-level and in the large volume limit, that is when the effects of worldsheet instantons can be neglected. Despite this approximation we have shown that certain D6-brane neutral fields, namely Wilson lines and special Lagrangian deformations do enter the tree-level flux potential in quite a similar fashion as the B-field axions do. More precisely, we have found that the way that they enter into the flux potential is dictated by a series of discrete shift symmetries, which amount to simultaneously perform loops in open string moduli space and shift the values of RR flux quanta. As these symmetries are also manifest at the level of the superpotential and they are ultimately related to how we define gauge invariant fluxes in 10d supergravity, we expect them to be present even after threshold corrections have been taken into account. 

The form of the open-closed scalar potential has non-trivial implications for the data of the 4d $\cn=1$ supergravity effective field theory. In particular it gives stringent constraints on how the appearance of open string moduli modifies the well-known K\"ahler potential for closed string modes. In this respect we have found several general features that the open-closed type IIA K\"ahler potential must satisfy at this level of approximation. First $e^K$ must be a real homogeneous function of degree seven on the imaginary part of the 4d chiral fields, implying a continuous shift symmetry for the D6-brane Wilson lines. This is a rather strong result but nevertheless in total agreement with the uplift of these compactifications to M-theory in $G_2$ manifolds \cite{Beasley:2002db}. Second we have found that in the presence of open string modes, the moduli spaces of complex structure and K\"ahler deformations no longer factorise, a result that seems to be mostly overlooked in the Calabi-Yau literature. Nonetheless, this is again in agreement with the well-known cases of type II compactifications in toroidal orientifolds. Notice that if we backreact D-branes sources the resulting warping effects are also expected to break such a factorisation \cite{Martucci:2009sf}, so it would be nice to see if these two effects are actually related. 

These general results for the K\"ahler potential are directly related to how closed string 4d holomorphic variables are redefined in the presence of D-branes degrees of freedom. Although our redefinitions differ from the previous proposals in the Calabi-Yau literature, in terms of K\"ahler potential modifications in D6-brane models they reproduce the proposal made in \cite{Grimm:2011dx} to embed Hitchin's functionals into the K\"ahler potential for complex structure moduli. This form of the open-closed K\"ahler potential will have important implications for models of large field inflation involving D-brane moduli, as will be discussed in \cite{backreaction}.

There are a number of directions in which our analysis can be generalised and that would be interesting to explore in the future. For instance, in our type IIA analysis we have only considered the scalar potential and superpotential generated by RR fluxes. While this is sufficient for the scope of this work, it would be important to also include NS fluxes in order to incorporate open string moduli in type IIA models of moduli stabilisation \cite{DeWolfe:2005uu,Camara:2005dc,Palti:2008mg}. In particular it would be interesting to analyse the interplay of two different effects of such NS fluxes: how they modify the closed string scalar potential and how they generate a discretum of D-brane positions \cite{Gomis:2005wc}. In a similar spirit, it would be interesting to generalise our type IIA analysis to non-K\"ahler flux compactifications. Since in these more general backgrounds the D6-brane deformations are also determined by the number non-trivial one-cycles of the wrapped three-cycle \cite{Marchesano:2006ns,Koerber:2006hh}, one may again consider applying Hitchin's functionals to describe the open string K\"ahler metrics. Furthermore, it would be interesting to analyse the different corrections that will modify the scalar potential analysed here. There would include threshold corrections to the K\"ahler potential \cite{Berg:2004ek,Berg:2005ja,Berg:2005yu,Haack:2008yb,Berg:2014ama}, worldsheet instanton corrections to the type IIA superpotential and D-brane instanton corrections. In particular, it would be nice to incorporate the latter directly into the three-form derivation of the scalar potential of section \ref{s:3forms}, following the recent proposal in \cite{Garcia-Valdecasas:2016voz}.

Finally, while our analysis has been restricted to non-chiral D-brane fields it would be interesting to analyse the consequences of our results for more realistic 4d models in which chiral matter arises from D-branes intersections. In this respect notice that  throughout our discussion a key role has been played by the discrete shifts in open string moduli space, and in particular those which leave invariant the open-closed superpotential. Remarkably, the same kind of shifts are the ones generating discrete flavour symmetries in simple orientifold models \cite{Marchesano:2013ega}. Therefore it would be interesting to see if there are semi-realistic string models in which the structure of the moduli stabilisation potential is directly related to the flavour structure of the D-brane sector.

\bigskip

\bigskip

\centerline{\bf \large Acknowledgments}

\bigskip

We would like to thank Sjoerd Bielleman, Luis Ib\'a\~nez, Michael Haack, Luca Martucci, Diego Regalado  and Irene Valenzuela for useful discussions. 
This work has been partially supported by the grants FPA2012-32828 and FPA2015-65480-P from  MINECO, SEV-2012-0249 of the ``Centro de Excelencia Severo Ochoa" Programme, and the ERC Advanced Grant SPLE under contract ERC-2012-ADG-20120216-320421. F.C. is supported through a fellowship of the international programme ``La Caixa-Severo Ochoa" and G.Z. is supported through a grant from ``Campus Excelencia Internacional UAM+CSIC". F. M. and G. Z. would like to thank UW-Madison for hospitality during completion of this work.

\clearpage

\appendix


\section{Details on the K\"ahler metrics}
\label{ap:details}

The K\"ahler potential that describes type IIA orientifold compactifications is given by
\be
K\, =\, K_K + K_Q \, =\, -{\rm log} \left(\cg_K\cg_Q^2\right)
\label{ap:K}
\ee
where 
\bea
\label{ap:KK}
K_K &  = & -{\rm log} \left(\frac{i}{6} \CK_{abc} (T^a - \bar{T}^a)(T^b - \bar{T}^b)(T^c - \bar{T}^c) \right)  = -{\rm log} \left(\cg_K\right) \\
K_Q &  = & -2\,{\rm log} \left(\frac{1}{4} \left[ \re(C\CF_\Lam)\im(CX^\Lam) - \re(CX^K)\im(C\CF_K) \right]  \right)  = -2{\rm log} \left(\cg_Q\right)
\label{ap:KQ}
\eea
with the definitions made in section \ref{s:typeIIA}. Due to the orientifold geometry the holomorphic three-form of the Calabi-Yau $\CM_6$ takes the form
\be
C\Om \, =\, \re(CX^K) \a_K + i \im(CX^\Lam)\a_\Lam -\re(C\CF_\Lam)\b^\Lam - i\im(C\CF_K)\b^K
\ee
where each of the coefficients are functions of the real parameters $n'^{K}$ and $u'_\Lam$ that define the complex structure of $\CM_6$. Following \cite{Candelas:1990pi} one can apply the equality
\be
\int_{\CM_6} \Om \wedge \p_{n'^{K}} {\Om}\, =\, \int_{\CM_6} \Om \wedge \p_{u'_\Lam} {\Om}\, =\, 0
\ee
in the above expression for $C\Omega$ to show that
\be
n'^{K} \p_{n'^{K}} \cg_Q + u'_\Lam \p_{u'_\Lam} \cg_Q = 2 \cg_Q
\ee
which means that $\cg_Q$ is a homogeneous function of degree two on the variables $n'^{K}$, $u'_\Lam$
\be
\cg_Q(\lam n'^{K}, \lam u'_\Lam)\, =\, \lam^2 \cg_Q(n'^{K}, u'_\Lam)
\ee
In addition, it is easy to see that $\cg_K$ is homogeneous of degree three on the variables $t^a = \im T^a$. Therefore the real function $\cg_K\cg_Q^2$ that appears in (\ref{ap:K}) is homogeneous of degree seven on the variables $\{t^a, n'^{K}, u'_\Lam\}$. 

As discussed in section \ref{s:kahler}, in order to introduce the open string moduli $\Phi^i$ into the K\"ahler potential one needs to express $n^{\prime K}$ and $u_\Lam^{\prime}$ in terms of new variables that depend on the open string position moduli. Recall that such moduli are described in terms of the functions $f_{\a \, a}^i$ as
\be
\phi^i_\a \, \equiv\, \im\, \Phi^i_\a\, =\, - t^a f^i_{\a\,  a} \, = \, t^a \frac{d \phi^i_\a}{d t^a}
\ee
where in the last equality we have used the definition (\ref{deffch}) and that $\om_a =  d\, \im\, J_c/dt^a$. If we see $\phi_a^i$ as functions of $t^a$ and $\varphi^j_\a$, the above relation  means that $\phi_a^i$ is homogeneous function of degree one on the K\"ahler moduli $t^a$, or in other words that we have the following scaling behaviour
\be
t^a \, \raw\, \lam t^a   \qquad \text{and} \qquad \phi_\a^k\,\raw\, \lam \phi_\a^k
\label{scalingt}
\ee
for the scaling of open string moduli in terms of K\"ahler moduli. As a direct consequence $f^i_{\a\,  a} \equiv f^i_{\a\,  a}(t^b, \varphi^j_\a) $ are homogenous functions of zero degree on $t^a$ or, if we see them as $f^i_{\a\,  a} \equiv f^i_{\a\,  a}(t^b, \phi^j_\a)$, they should be invariant under the simultaneous rescaling (\ref{scalingt}). These statements are equivalent to
\be
t^a \frac{d}{dt^a} f^i_{\a\,  b}\, =\, t^a \left( \p_{t^a} +  \frac{\p\phi^j_\a}{\p t^a} \p_{\phi^j_\a}\right)f^i_{\a\,  b}\, =\, \left( t^a\p_{t^a} +  \phi^j_\a \p_{\phi^j_\a}\right)f^i_{\a\,  b}\, =\ 0
\ee
Finally, as pointed out in the main text $f^i_{\a\, a}$ may also depend on the complex structure moduli $n'^K$ and $u'_\Lam$. Now because the harmonic two-forms $\om_a$ are invariant under an overall rescaling of the holomorphic three-form $\Omega$, they can only depends on quotients of their periods, and so the same should be true for $f_{\a \, a}^i$. Therefore these functions should also be  invariant under the rescaling
\be
n'^K \, \raw\, \lam' n'^K   \qquad \text{and} \qquad u'_\Lam \,\raw\, \lam' u'_\Lam
\label{scalingnu}
\ee
with $\lam'$ independent from $\lam$ in (\ref{scalingt}). Finally, a similar reasoning can be applied to the functions $g_{\a\, i}^{K}$ and $g_{\a\, \Lam\, i}$, defined in (\ref{defgs}). Indeed, from such a chain integral expression one can argue that these functions should also be invariant under (\ref{scalingt}) and (\ref{scalingnu}) separately, and in particular homogeneous functions of zero degree on the variables $\{t^a, n'^{K}, u'_\Lam, \phi^j_\a\}$.

These observations are relevant for the redefinition of holomorphic variables proposed in section \ref{s:kahler}, which imply that we must perform the following replacement in $\cg_Q$
\be
n^{\prime K} \, \raw \,  n^{K} + \oh t^a \sum_\a {\bf H}^K_{\a\, a} \qquad
u_\Lam^{\prime} \, \raw \, u_\Lam - \oh t^a \sum_\a  {\bf H}_{\a \, \Lam\, a}  
\label{ap:nuprime}
\ee
with $n^K =  \im\, N^{K}$, $u_\Lam = \im\, U_\Lam$ the imaginary parts of the new holomorphic variables. Here ${\bf H}^K_{\a\, a}$ and ${\bf H}_{\a \, \Lam\, a}$ must satisfy (\ref{defH}) and (\ref{redefUlam}), or equivalently
\be
\p_{\phi_\b^j} (t^a {\bf H}^K_{\a\, a})\, =\, - g_{\a\, i}^K \d_{\a\b} \qquad \qquad \p_{\phi_\b^j} (t^a {\bf H}_{\a \, \Lam\, a}) \, =\, - g_{\a \, \Lam\, i}\d_{\a\b}
\label{ap:defH}
\ee
so we can see them as functions of the variables $\{t^a, n'^{K}, u'_\Lam, \phi^j_\a\}$. Since $g_{\a\, i}^{K}$ and $g_{\a\, \Lam\, i}$ are homogeneous of zero degree on these variables, the same can be assumed for ${\bf H}^K_{\a\, a}$, ${\bf H}_{\a \, \Lam\, a}$. Finally, by recursively performing the replacement (\ref{ap:nuprime}) we can see ${\bf H}^K_{\a\, a}$, ${\bf H}_{\a \, \Lam\, a}$ as homogeneous functions of zero degree on $\{t^a, n^K, u_\Lam, \phi_\a^k\}$.

We then see that the rhs of (\ref{ap:nuprime})  are homogeneous functions of degree one on the real fields $\{t^a, n^K, u_\Lam, \phi_\a^k\}$ on which the K\"ahler potential depends. As a consequence, when we perform such replacements we obtain that $\cg_Q$ remains a homogeneous function of degree two in  the new fields, namely
\be
\cg_Q(\lam t^a, \lam n^K, \lam u_\Lam\, \lam \phi_\a^k)\, =\, \lam^2 \cg_Q (t^a, n^K, u_\Lam\, \phi_\a^k)
\ee
Finally, we have that 
\be
\cg (\psi^\a) \, =\, \cg_K\cg_Q^2
\ee
is homogeneous of degree seven on the whole set of fields $\psi^\a \equiv \{t^a, n^K, u_\Lam, \phi^k\}$, where for simplicity we have absorbed the D6-brane index $\a$ into the index $i$.

From this simple observation several useful relation can be derived \cite{Ferrara:1994kg}. For instance
\be
K^{\a\bar{\b}}K_{\bar{\b}}\, =\, - (\Psi^\a- \bar{\Psi}^\a)\, \equiv\, -2i \psi^\a
\label{ap:sprel}
\ee
with $\Psi^a$ any of the complex fields of the compactification. To see this we first we rewrite this relation as
\be
-2i K_{\bar{\a}\b} \psi^\b\, =\, K_{\bar{\a}}
\label{ap:sprel2}
\ee
which is easier to check. Then we compute
\bea
K_{\bar{\a}} & = & \frac{1}{2i} \frac{\p_\a \cg}{\cg}\\
K_{\bar{\a}\b} & = & -\frac{1}{4} \left(\frac{\p_\a\p_\b\cg}{\cg} - \frac{\p_\a\cg\p_\b\cg}{\cg^2} \right)
\eea
and then use the relations
\be
\psi^\b\p_\b\cg \, =\, 7 \cg \quad \quad \psi^\b\p_\a\p_\b\cg \, =\, 6 \p_\a \cg
\label{ap:homorel}
\ee
that arise from the homogeneity of $\cg$ to prove (\ref{ap:sprel2}). Moreover, using the first identity in (\ref{ap:homorel}) again, one can show that the no-scale relation
\be
K^{\a\bar{\b}}K_\a K_{\bar{\b}}\, =\,  7
\label{ap:noscale}
\ee
follows automatically. Finally, one can use these relations to get a simple  expression for the inverse K\"ahler metric
\be
K^{\bar{\a}\b} \, =\, \frac{2}{3}  \psi^\a\psi^\b - 4 \cg\cg^{\a\b}
\ee
where $\cg^{\a\b}$ is the inverse of $\p_\a\p_\b \cg$.  For a recent general discussion on no-scale K\"ahler potentials based on homogeneous functions an their generalisation see \cite{Ciupke:2015ora}.

\subsubsection*{Relations for the inverse metric}

We would now like to discuss several identities relating the inverse K\"ahler metric components which are important for the computations of section \ref{s:spot}, and see how they may arise from the above K\"ahler potential. For simplicity we will work in the symplectic basis defined above eq.(\ref{KQN}) so that $\cg_Q$ is a homogeneous function of degree two purely on the variables $n'^K$, which in turn depend on the holomorphic fields $N^K$, $T^a$ and $\Phi^i_\a$ and their conjugates through (\ref{ap:nuprime}). As this dependence can in general be quite involved, in here we will make the simplifying assumption that the functions $f^i_{\a\,  a}$ and $g_{\a\, i}^{K}$ only depend on $\varphi^j_\a$, which will allow us to carry the computations analytically.\footnote{This assumption is valid in simple cases like toroidal compactifications, and it should be a good approximation in the large volume and complex structure regions of the Calabi-Yau moduli space. Indeed, in general, $f^i_{\a\,  a}$ and $g_{\a\, i}^{K}$ are homogeneous functions of zero degree on $\{t^a, n'^{K}, u'_\Lam, \phi^j_\a\}$, invariant under (\ref{scalingt}) and (\ref{scalingnu}). As such they depend on $\varphi^j_\a$ and also on quotients of K\"ahler moduli $t^b/t^a$ and complex structure moduli $n'^K/n'^J$,  $u'_\Lam/u'_\Sig$. The dependence on these quotients is very mild for large values of these bulk fields, that is in the regions of large volume and complex structure, where the bulk harmonic forms $\omega^a$, $\beta^K$ do not vary significantly with respect to variations of the closed string moduli.}

Because $\varphi^j_\a \equiv \varphi^j_\a(\phi^i_\a, t^a)$, we may then see $f^i_{\a\,  a}$ and $g_{\a\, i}^{K}$ as functions of $\phi^i_\a$ and $t^a$, but nevertheless such that $d f^i_{\a\,  a}/dt^b = d g_{\a\, i}^{K}/dt^b = 0$. Given the definition (\ref{defH}) the same applies to ${\bf H}^K_{\a\, a}$, and so we have that
\be
\frac{d {\bf H}^K_{\a\, a}}{dt^b} \, =\, \left( \p_{t^b} +  \frac{\p\phi^j_\a}{\p t^b} \p_{\phi^j_\a}\right) {\bf H}^K_{\a\, a} \, =\, 0 \quad \Raw \quad \p_{t^b} (t^a {\bf H}^K_{\a\, a}) \, =\, {\bf H}^K_{\a\, b} -  f^i_{\a\, b} g_{\a\, i}^{K}
\label{ap:ptH}
\ee
which will be used later.

From the assumption that ${\bf H}_{\a\, i}^K$ does not depend on the complex structure moduli one can see that the K\"ahler metric can be written in the form
\be
{\bf K}\, =\, 
\begin{blockarray}{(ccc)}
\ \Id & 0 & 0 \ \\
\ {\bf T}^\dag & \Id & 0 \ \\
\ {\bf \Phi}^\dag & 0 & \Id \ \\
 \end{blockarray}
 \quad
\begin{blockarray}{(ccc)}
        \ {\bf N}& 0 & 0 \  \\
        \ 0 &  \BAmulticolumn{2}{c}{\multirow{2}{*}{\Large${\bf \Omega}$}}\\ 
         \ 0 \\
     \end{blockarray}
     \quad
\begin{blockarray}{(ccc)}
\ \Id & {\bf T} & {\bf \Phi} \ \\
\ 0 & \Id & 0 \ \\
\ 0 & 0 & \Id \ \\
   \end{blockarray}
\ee
where we have defined the matrices
\be
{\bf N}_{K\bar{L}}\, =\, \p_{N'^K}\p_{\bar{N}'^{{L}}} {\bf K_Q} \quad \quad {\bf T}^{\bar{L}}{}_{\bar{a}}\, =\, \p_{\bar{T}^{{a}}} \bar{N}'^L \quad \quad {\bf \Phi}^{\bar{L}}{}_{\bar{i}}\, =\, \p_{\bar{\Phi}^{{i}}} \bar{N}'^{{L}} 
\ee
and
\be
{\bf \Omega}\, =\, 
\left(
\begin{array}{cc}
{\bf A} & {\bf B} \\
{\bf C} & {\bf D}
\end{array}
\right)
\ee
with
\bea
{\bf A}_{a\bar{b}} & = & \p_{T^a}\p_{\bar{T}^b}{\bf K_K} +  (\p_{n'^K} {\bf K_Q}) \p_{T^a}\p_{\bar{T}^b} n'^{K}\\
{\bf B}_{a\bar{j}} & = & (\p_{n'^K} {\bf K_Q}) \p_{T^a}\p_{\bar{\Phi}^j} n'^{K}\\
{\bf D}_{i\bar{j}} & = & (\p_{n'^K} {\bf K_Q}) \p_{\Phi^i}\p_{\bar{\Phi}^j} n'^{K} 
\eea
and ${\bf C} = {\bf B}^\dag$. For simplicity we have absorbed the D6-brane index $\a$ into the index $i$ counting open string moduli. 

From this expression we find that the inverse metric is given by
\be
{\bf K}^{-1}\, =\, 
\begin{blockarray}{(ccc)}
\ \Id & -{\bf T} & -{\bf \Phi} \ \\
\ 0 & \Id & 0 \ \\
\ 0 & 0 & \Id \ \\
 \end{blockarray}
 \quad
\begin{blockarray}{(ccc)}
        \ {\bf N}^{-1}& 0 & 0 \  \\
        \ 0 &  \BAmulticolumn{2}{c}{\multirow{2}{*}{\Large${\bf \Omega}^{-1}$}}\\ 
         \ 0 \\
\end{blockarray}
 \quad
\begin{blockarray}{(ccc)}
\ \Id & 0 & 0 \ \\
\ -{\bf T}^\dag & \Id & 0\  \\
\ -{\bf \Phi}^\dag & 0 & \Id \ \\
\end{blockarray}
\ee
where
\be
{\bf \Omega}^{-1}\, =\,
\left(
\begin{array}{cc}
\Id & 0 \\
-{\bf D}^{-1}{\bf C} & \Id
\end{array}
\right)
\left(
\begin{array}{cc}
({\bf A} - {\bf B}{\bf D}^{-1}{\bf C})^{-1} & 0 \\
0 & {\bf D}^{-1}
\end{array}
\right)
\left(
\begin{array}{cc}
\Id & -{\bf B}{\bf D}^{-1} \\
0 & \Id
\end{array}
\right) \ .
\ee

From here we deduce the relation 
\be
{\bf K}^{\bar{a}i} +  {\bf K}^{\bar{a}b} {\bf B}_{b\bar{j}}  {\bf D}^{\bar{j}i} \, =\, 0
\label{interid}
\ee
and so the relation (\ref{keyrel}) can be rephrased as
\be
{\bf B}_{a\bar{j}} \, =\,  f_{a}^i  {\bf D}_{i\bar{j}}
\label{ap:keyrel}
\ee
or equivalently
\be
\im(C\CF_K) \left[ \p_{t^a} +  \frac{\p \phi^i}{\p t^b} \p_{\phi^i}\right] (\p_{\phi^j} n'^K) \,=\, \im(C\CF_K) \frac{d}{dt^a} (\p_{\phi^j} n'^K)\, =\, 0
\label{ap:keyrels}
\ee
where as before $t^a = \im\, T^a$ and $\phi^i = \im\, \Phi^i$. Using that $\p_{\phi^j} n'^K = -\oh g_{i}^K$  and the assumption $dg_{i}^K/dt^a = 0$ we recover the desired identity. 

From these expressions one also obtains that 
\bea
{\bf A}_{a\bar{b}} - {\bf B}_{a\bar{j}}{\bf D}^{\bar{j}{i}}{\bf C}_{i\bar{b}} & = &  \p_{T^a}\p_{\bar{T}^b}{\bf K_K} + (\p_{n'^K} {\bf K_Q}) \left( \p_{T^a}\p_{\bar{T}^b} - f_a^i  \p_{\Phi^i} \p_{\bar T^b} \right) n'^{K}\\ \nonumber
& = &  \p_{T^a}\p_{\bar{T}^b}{\bf K_K} + \frac{1}{4} (\p_{n'^K} {\bf K_Q})  \frac{d}{dt^a} (\p_{t^b} n'^K) \, =\,  \p_{T^a}\p_{\bar{T}^b}{\bf K_K} 
\eea
where in the last equality we have used (\ref{ap:ptH}). Hence the inverse K\"ahler metric for the K\"ahler moduli is exactly the same as in the absence of open string degrees of freedom. More precisely we have that
\be
K^{a\bar{b}}\, =\, 2 t^at^b -\frac{2}{3}\CK\CK^{ab}
\label{ap:invKab}
\ee
where we have defined $\CK$ as in (\ref{Ks}) and $\CK^{ab}$ is the inverse of $\CK_{ab}$ in there. That is, we have the same inverse metric as we would have if there were no open string moduli. 
Finally, applying \eqref{ap:keyrel} we have that
\be
K^{i \bar \jmath} -  K^{a \bar b}f_a^if_b^j\, =\, {\bf D}^{i \bar \jmath}
\label{ap:invKij}
\ee
where ${\bf D}_{i \bar \jmath}$ is given by 
\be
{\bf D}_{i \bar \jmath}\, =\, \frac{1}{4} \p_{n'^K} \mathbf{K_Q} \p_{\phi^i} \,\p_{\phi^j}n'^K\, =\, - \p_{n'^K} \mathbf{K_Q}\, \frac{1}{8} \p_{\phi^i}  g^K{}_j 
\label{ap:Dij}
\ee
Now in order to match the DBI potential from section \ref{s:spot} we need that
\be
{\bf D}^{i \bar \jmath}\, =\, G^{ij}_{\rm D6}\, =\, 8  \hat{V}_6\, e^{\phi/4}\, l_s^{-3} \int_{\Pi_\a} \rho^i \wedge * \,\rho^j
\label{ap:invgD6}
\ee
or equivalently
\bea \nonumber
{\bf D}_{i \bar \jmath} & = & G_{ij}^{\rm D6} \, = \, \frac{e^{-\phi/4}}{8 \hat{V}_6} l_s^{-3} \int_{\Pi_\a} \zeta_i \wedge *\, \zeta_j \\
& = & -  \frac{1}{8 \hat{V}_6} \left( l_s^{-3} \int_{\Pi_\a} \iota_{X_k} J \wedge \rho^i\right)^{-1} \left( l_s^{-3} \int_{\Pi_\a} \iota_{X^k} \im\, (C\Om) \wedge \zeta_j\right) \\
\nonumber
&=& \frac{1}{8 \hat{V}_6}   \im(C\CF_K)  (\CQ^K)_{jk} \left[ (t^a\eta_a)^{-1}\right]^k_i
\label{ap:gD6}
\eea
where we have used that $e^{-\phi/4}  [\iota_{X^j} J]_{\Pi_\a}  = - *_3 [\iota_{X^j} \im\, (C \Om)]_{\Pi_\a}$  \cite{McLean}. Comparing with (\ref{ap:Dij}) this implies that
\be
\p_{\phi^i}  g^K{}_j  \, =\, - (\CQ^K)_{jk} \left[ (t^a\eta_a)^{-1}\right]^k_i  \quad \iff \quad \p_{\varphi^i}  g^K{}_j\, =\, (\CQ^K)_{ji}
\ee
in agreement with (\ref{pgQ}). In fact, notice that we will also reproduce the same set of results if in (\ref{ap:defH}) we shift $g^K{}_j$ by a constant, which means that the $t^a {\bf H}^K_a$ are only  determined up to a linear function on $\phi^j$.

\section{The closed string scalar potential}
\label{ap:dimred}

The literature already offers several approaches to obtain the scalar potential for the K\"ahler moduli in presence of background RR fluxes through dimensional reduction of (massive) type IIA supergravity, both excluding~\cite{Louis:2002ny,Grimm:2004ua,Bielleman:2015ina} and including \cite{Grimm:2011dx,Kerstan:2011dy} couplings to D-branes. In the absence of D6-branes, the scalar potential arises from the kinetic terms of the RR field strengths upon dimensional reduction of the standard formulation of (massive) type II supergravity. In the presence of D6-branes, the road to follow for the dimensional reduction passes through the democratic formulation of massive type IIA supergravity, where aside from the RR-potentials $C_1$ and $C_3$ and Romans mass parameter $m$ also the dual RR-potentials $C_5$, $C_7$ and $C_9$ are taken into consideration and the Hodge duality relations are usually imposed by hand. In this formulation the scalar potential for the K\"ahler moduli emerges upon dualisation of the four-dimensional four-forms associated to the dimensional reduction of the RR-field strengths in favour of the RR-flux quanta. 

This appendix offers a first principle approach to deduce a four-dimensional mother action allowing for the dualisation of the four-forms also in the presence of D6-branes, while taking into account the considerations about Dirac quantisation around equation~(\ref{Diracq}). As advertised in~\cite{Bergshoeff:2001pv,Gukov:2002iq,Villadoro:2005cu} background geometries including closed string fluxes and D-branes can be better approached from the A-basis,  instead of the C-basis. For this reason we choose to start from the following mother action:
\begin{align} \label{Eq:MAGPF}
{\cal S}^{\rm mother}_{RR} =&  \frac{1}{2 \kappa_{4}^2 l_s^6}\int \left[ -\sum_{p=0}^5 \frac{1}{4} G_{2p} \wedge \star G_{2p} + \frac{1}{2} A_9 \wedge d G_0 - \frac{1}{2} A_7 \wedge d \left(G_2 - G_0 B\right) \right. \nonumber \\
 & \hspace{0.8in}+ \frac{1}{2} A_5 \wedge d\left(G_4 - G_2 \wedge B + \frac{G_0}{2} B^2\right) \\
& \hspace{0.8in}  - \frac{1}{2} A_3 \wedge d\left(G_6 - G_4 \wedge B + \frac{1}{2} G_2 \wedge B^2 -  \frac{G_0}{3!} B^3\right)  \nonumber \\
 &\hspace{0.8in} \left.  +  \frac{1}{2} A_1 \wedge d\left(G_8 - G_6 \wedge B + \frac{1}{2} G_4 \wedge B^2 -  \frac{1}{3!} G_2\wedge B^3 + \frac{G_0}{4!} B^4 \right) \right] \nonumber, 
\end{align}  
where the potentials ${\bf A}$ are playing the r\^ole of Lagrange multipliers imposing the Bianchi identities for the field strengths in the A-basis. The solutions of the Bianchi identities correspond to the field strengths $G_{2p}$ given in terms of the RR-potentials in the A-basis and constant fluxes as indicated in equation~(\ref{BIG}). Using the basis of harmonic forms on ${\cal M}_6$ introduced in section~\ref{s:3forms} for a background with RR-fluxes only, the dimensional reduction of the field strengths contains  the flux quanta~(\ref{RRfluxes}):
\begin{align}
G_0 &= l_s^{-1} m, \nonumber \\
G_2 & = l_s^{-1} \left( m^a  + b^a m\right)  \omega_a + \ldots, \nonumber  \\
G_4 &= l_s^{-1} \left((e_a + {\cal K}_{abc} m^b b^c + \frac{m}{2}  {\cal K}_{abc} b^b b^c \right) \tilde \omega^a  +  D_4^0 + \ldots, \\
G_6 & =  l_s^{-1} \left(e_0 + e_a b^a + \frac{1}{2}{\cal K}_{abc} m^a b^b b^c + \frac{m}{3!}  {\cal K}_{abc} b^a b^b b^c \right)  \omega_6 + \left(D_4^a + b^a D_4^0\right) \wedge \omega_a +\ldots, \nonumber \\
G_8 &= \left( \tilde D_{4a} + {\cal K}_{abc} b^b D^c_4 + \frac{1}{2} {\cal K}_{abc} b^b b^c D_4^0 \right) \wedge \tilde \omega^a + \ldots, \nonumber \\
G_{10} &= \left(\tilde D_4 + \tilde D_{4a }b^a + \frac{1}{2} {\cal K}_{abc} b^a b^b D^{c}_4 + \frac{1}{3!} {\cal K}_{abc} b^a b^b b^c D_4^0   \right) \wedge \omega_6 + \ldots . \nonumber
 \end{align} 
Inserting the expansion back into the mother action~(\ref{Eq:MAGPF}), converting to the Einstein-frame by virtue of a rescaling of the ten-dimensional metric $G_{MN} \rightarrow e^{\frac{\phi}{2}} G_{MN}$ and rescaling the four-dimensional metric $g_{\mu \nu}\rightarrow\frac{g_{\mu \nu}}{ \hat V_6/2}$, we obtain the following four-dimensional effective action (exploiting the notations of section~\ref{s:3forms}):
\begin{align}
4\kappa_4^2 {\cal L}^{\rm mother}_{4d} =& - \frac{1}{4 }  \frac{4}{\hat V_6} e^{\frac{5\phi}{2}}\tilde \rho^2   - \frac{1}{4}e^{\frac{5\phi}{2}} \frac{16}{\hat V_6}   g_{ab}  \tilde \rho^a \tilde \rho^b   - \frac{1}{4} e^{-\frac{\phi}{2}}  \frac{g^{ab}}{\hat V_6^3} \rho_a \rho_b - \frac{1}{4} e^{-\frac{\phi}{2}}  \frac{4}{\hat V_6^3} \rho_0^2 \nonumber \\
& - \frac{1}{4} \frac{\hat V_6}{4} e^{-\frac{5\phi}{2}} \left(  \tilde D_4 + \tilde D_{4a }b^a + \frac{1}{2} {\cal K}_{amn} b^a b^m D^{n}_4 + \frac{1}{3!} {\cal K}_{amn} b^a b^m b^n D_4^0   \right)   \nonumber\\  
& \qquad \qquad  \wedge *\left(\tilde D_4 + \tilde D_{4b }b^b + \frac{1}{2} {\cal K}_{brs} b^b b^r D^{s}_4 + \frac{1}{3!} {\cal K}_{brs} b^b b^r b^s D_4^0 \right) \nonumber\\ 
 &  - \frac{1}{4}  e^{-\frac{5\phi}{2}} \frac{\hat V_6 g^{ab}}{ 16} \left( \tilde D_{4a} + {\cal K}_{amn} b^m D^n_4 + \frac{1}{2} {\cal K}_{amn} b^m b^n D_4^0  \right) \nonumber \\
 & \qquad \qquad   \wedge * \left( \tilde D_{4b} + {\cal K}_{brs} b^r D^s_4 + \frac{1}{2} {\cal K}_{brs} b^r b^s D_4^0  \right) \nonumber \\
 &  - \frac{1}{4} e^{\frac{\phi}{2}}  g_{ab} \hat V_6^3  \left( D_4^a + b^a D_4^0 \right)  \wedge * \left( D_4^b + b^b D_4^0 \right)   - \frac{1}{4} e^{\frac{\phi}{2}}  \frac{\hat V_6^3}{4} D_4^0 \wedge *  D_4^0 \nonumber \\
& - \frac{1}{2} l_s^{-1} m \tilde D_4 +  \frac{1}{2} l_s^{-1} m^a \tilde D_{4a}     - \frac{1}{2}  l_s^{-1} e_a D_4^a + \frac{1}{2} l_s^{-1} e_0 D_4^0.
\end{align}
By integrating out the four-forms in the order $\tilde D_{4}\rightarrow \tilde D_{4a}  \rightarrow D_4^a \rightarrow D_4^0$ in favour of the flux quanta, the usual four-dimensional effective scalar potential for the K\"ahler moduli arises as given in equations~(\ref{VRR}) and~(\ref{expotRR}).
Alternatively, one can rotate the four-forms $(D_4^0,D_4^a,\tilde D_{4a},\tilde D_{4})$ into the four-forms $(F_4^0,F_4^a, \tilde F_{4a}, \tilde{F}_4)$, in which case the four-dimensional mother action reads:
\begin{align}
4\kappa_4^2{\cal L}^{\rm mother}_{4d} =& - \frac{1}{4 } e^{\frac{5\phi}{2}}   \frac{4}{\hat V_6} \tilde \rho^2   - \frac{1}{4 } e^{\frac{5\phi}{2}}  \frac{16}{\hat V_6}   g_{ab}  \tilde \rho^a \tilde \rho^b   - \frac{1}{4} e^{-\frac{\phi}{2}}  \frac{g^{ab}}{\hat V_6^3} \rho_a \rho_b - \frac{1}{4} e^{-\frac{\phi}{2}}  \frac{4}{\hat V_6^3} \rho_0^2  \nonumber \\
& - \frac{1}{4} e^{-\frac{5\phi}{2}} \frac{\hat V_6}{4} \tilde F_4 \wedge * \tilde F_4  - \frac{1}{4} e^{-\frac{5\phi}{2}}  \frac{\hat V_6 g^{ab}}{ 16} \tilde F_{4a} \wedge * \tilde F_{4b} \nonumber\\& - \frac{1}{4} e^{\frac{\phi}{2}} g_{ab} \hat V_6^3 F^{a}_4\wedge * F^b_4 - \frac{1}{4}  e^{\frac{\phi}{2}}  \frac{\hat V_6^3}{4} F_4^0 \wedge *  F_4^0\nonumber \\
& + \frac{1}{2} \tilde \rho \tilde F_4 + \frac{1}{2} \tilde \rho^a \tilde F_{4a} + \frac{1}{2}  \rho_a F^{a}_4 + \frac{1}{2} \rho_0 F_4^0
\end{align}
This change of four-form base boils down to a rewriting of the mother action in the C-basis, where we recombined the flux quanta into the Lagrange-multipliers $(\rho_0,\rho_a,\tilde \rho^a,\tilde \rho)$  as in equation (\ref{rhos}) for simplicity. Eliminating the four-forms $(F_4^0,F_4^a, \tilde F_{4a}, \tilde{F}_4)$ through their equations of motion: 
\begin{equation}\label{Eq:EOMFrho}
\begin{array}{rcl}
\tilde \rho &=&  e^{-\frac{5\phi}{2}}\frac{\hat V_6}{4} * \tilde F_4  ,\\
 \tilde \rho^a& = & e^{-\frac{5\phi}{2}} \frac{\hat V_6 g^{ab}}{ 16} * \tilde F_{4b} , \\
 \rho_a &=& e^{\frac{\phi}{2}} g_{ab} \hat V_6^3 * F^{b}_4, \\
 \rho_0 & =& e^{\frac{\phi}{2}} \frac{\hat V_6^3}{4} * F_4^0,
\end{array}
\end{equation}
yields the scalar potential from the RR fluxes as in equations~(\ref{VRR}) and~(\ref{expotRR}). Note that a mother action is not unique and a (classical) theory expressed in a particular set of degrees of freedom can arise from two different mother action; the only requisite however is that mother actions reproduce identical equations of motion (and Bianchi identities). It is trivial to see that the mother action (\ref{S4form}) reproduces the same equations of motion in~(\ref{Eq:EOMFrho}) for the four-forms $(F_4^0,F_4^a, \tilde F_{4a}, \tilde{F}_4)$ upon the identification $e^{K} = (8 e^{\phi/2} \hat V_6^3)^{-1}$. 

The virtue of the mother action~(\ref{S4form}) lies in the straightforward generalisation for backgrounds with RR-fluxes and D6-branes, whose Chern-Simons coupling to the bulk degrees of freedom can be captured by a shift of the flux quanta~(\ref{shiftedflux}). Including the contribution of a single D6-brane in the mother action and following the procedures outlined in section~\ref{s:3forms} we obtain the RR scalar potential in which K\"ahler moduli and open string moduli mix:
\begin{equation}
 \label{expotRROpen}
 \begin{array}{rcl}
 \kappa_4^2 V_{RR} & = & \frac{e^{-\frac{\phi}{2}} }{2 l_s^2 \hat{V}_6^3} \left( e_0 + b^a e_a  + \oh \CK_{abc}m^ab^bb^c + \frac{m}{6} \CK_{abc}b^ab^bb^c \right. \\
 &&\hspace{0.3in}\left. +  n_{F\, i}  \theta^i  -  n_{a \, i}\th^i b^a - n_{F\, i} f_a^i  b^a   +  n_{ai} f_c^i  b^a b^c 
 \right)^2\\
 & + &\frac{e^{-\frac{\phi}{2}} }{8 l_s^2 \hat{V}_6^3}  g^{ab} \left( e_a +\CK_{acd}m^cb^d + \frac{m}{2} \CK_{acd}b^cb^d  -  n_{a \, i}\th^i     - n_{F\, i} f_a^i  +   \CK_{abc} q^b b^c    \right)  \\ 
 &&\hspace{0.3in} \times \left( e_b +\CK_{bef}m^eb^f + \frac{m}{2} \CK_{bef}b^eb^f   -  n_{b \, k}\th^k  - n_{F\, k} f_e^k   +  \CK_{bef}q^e  b^f     \right) \\
 & + & \frac{2 e^{\frac{5\phi}{2}}}{l_s^2 \hat{V}_6}  g_{ab} \left( m^a + mb^a + q^a  \right)  \left( m^b + mb^b +  q^b \right)  \, +\, \frac{e^{\frac{5\phi}{2}}}{2l_s^2 \hat{V}_6} m^2
 \end{array}
\end{equation}

\section{A toroidal orbifold example}
\label{ap:texample}

To clarify the geometric origin of the open-closed superpotential from section \ref{s:supo} and the emergence of the open string moduli in the K\"ahler potential, we consider an explicit realisation on the orientifold $(T^2\times K3)/\Omega_p (-)^{F_L} {\cal R}$. $K3$ is considered in the orbifold limit $T^4/\Z_2$~\cite{Blumenhagen:2002wn} inheriting bulk two-cycles from the covering four-torus, while the $\Z_2$ action implies the existence of 16 exceptional two-cycles $e_{ij}$ stuck at the $\Z_2$-fixed points $i,j \in \{1,2,3,4\}$. For simplicity, we choose the $T^4$ to be factorisable and choose the root-lattice of $SU(2) \times SU(2)$ for each separate two-torus $T^{2}$. Factorisable bulk three-cycles $\Pi^{\rm bulk}_\alpha$ on $T^2_{(1)}\times T^4/\Z_2$ are expressed as linear combinations of basis three-cycles $(\pi_{\alpha_k},\pi_{\beta^k})$, which represent Poincar\'e dual three-cycles to the symplectic basis ($\alpha_k, \beta^k)$ of (bulk) (2,1)-forms in $H^3(T^2\times K3)$:
 \begin{equation}
\begin{array}{l@{\hspace{0.4in}}l}
\alpha_0  = dx^1 \wedge dx^2 \wedge dx^3, & \beta^0  = - dy^1 \wedge dy^2 \wedge dy^3,\\
\alpha_1  = dx^1 \wedge dy^2 \wedge dy^3, & \beta^1  = - dy^1 \wedge dx^2 \wedge dx^3,\\
\alpha_2  = dy^1 \wedge dx^2 \wedge dy^3, & \beta^2  = - dx^1 \wedge dy^2 \wedge dx^3,\\
\alpha_3  = dy^1 \wedge dy^2 \wedge dx^3, & \beta^3  = - dx^1 \wedge dx^2 \wedge dy^3.\\
\end{array}
\end{equation}
These generators of bulk three-cycles arise by considering $\Z_2$-invariant product cycles of the one-cycle $\pi_1$ or $\pi_2$ on $T_{(1)}^2$ and factorisable bulk two-cycles on $T^4/\Z_2$:
\begin{equation}
\begin{array}{c@{\hspace{0.4in}}c}
\pi_{\beta^0} = \pi_{1} \otimes  \pi_{3} \otimes \pi_{5} ,& \pi_{\alpha_0} = - \pi_{2} \otimes  \pi_{4} \otimes \pi_{6},\\
\pi_{\beta^1} = \pi_{1} \otimes  \pi_{4} \otimes \pi_{6} ,& \pi_{\alpha_1} = - \pi_{2} \otimes  \pi_{3} \otimes \pi_{5} ,\\
\pi_{\beta^2} = \pi_{2} \otimes  \pi_{3} \otimes \pi_{6}  ,& \pi_{\alpha_2} = - \pi_{1} \otimes  \pi_{4} \otimes \pi_{5} ,\\
\pi_{\beta^3} = \pi_{2} \otimes  \pi_{4} \otimes \pi_{5} ,& \pi_{\alpha_3} = - \pi_{1} \otimes  \pi_{3} \otimes \pi_{6}.\\
\end{array}
\end{equation}
The generators $(\pi_{\alpha_k}, \pi_{\beta^k})$ in $H_3 (T^2\times K3,\Z)$ form a symplectic basis of bulk three-cycles with ${\cal R}$-even three-cycles $\pi_{\beta^k}$ and ${\cal R}$-odd three-cycles $\pi_{\alpha_k}$, and were chosen in such a way that all complex structure moduli are of the $N^k$-kind as discussed in section~\ref{s:typeIIA}. Exceptional three-cycles $\Pi_\alpha^{\rm ex}$ can be expressed in terms of the generators $(\varepsilon_{ij} , \tilde \varepsilon_{ij} )$, constructed as $\Z_2$-invariant direct products of the one-cycle $\pi_1$ or $\pi_2$ on $T_{(1)}^2$ and an exceptional divisor $e_{ij}$ on $T^4/\Z_2$:
\begin{equation}
\varepsilon_{ij} = \pi_1 \otimes e_{ij}, \hspace{0.4in} \tilde \varepsilon_{ij} = \pi_2 \otimes e_{ij}, \hspace{0.4in} i,j \in\{1,2,3,4\}.
\end{equation}
Under the orientifold projection, the exceptional divisors pick up a minus sign,\footnote{This minus sign can be traced back to the involution operation $J \stackrel{{\cal R}}{\longrightarrow} -J$, translated to  the de Rahm dual $(1,1)$-forms situated at the blown-up singularities.} more explicitly ${\cal R}(e_{ij})=-e_{ij}$, such that exceptional three-cycles can be decomposed into ${\cal R}$-even exceptional three-cycles $\tilde \varepsilon_{ij}$ and ${\cal R}$-odd exceptional three-cycles $\varepsilon_{ij}$. The lattice of three-cycles generated by $\{\pi_{\alpha_k}, \pi_{\beta^k}, \varepsilon_{ij}, \tilde \varepsilon_{ij} \}$ has to be supplemented with fractional three-cycles:
\begin{equation}
\Pi_\alpha^{\rm frac} = \frac{1}{2} \Pi_\alpha^{\rm bulk} + \frac{1}{2} \Pi_\alpha^{\rm ex},
\end{equation}
to obtain the full lattice of three-cycles $H_3(T^2\times K3, \Z)$.

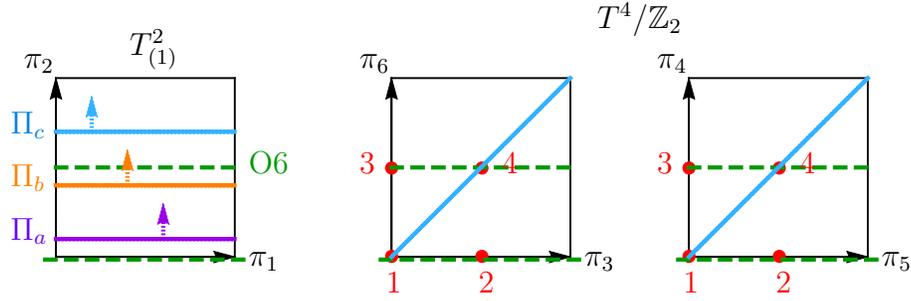
\begin{figure}[h]
\begin{center}
\vspace*{0.2in}
\begin{tabular}{c@{\hspace{0.6in}}c@{\hspace{0.4in}}c}
\includegraphics[width=2.8cm]{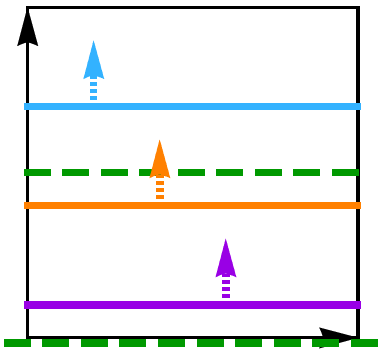} \begin{picture}(0,0) \put(-50,80){$T_{(1)}^2$} \put(-95,10){{\color{mypurple}$\Pi_a$}} \put(-95,30){{\color{myorange}$\Pi_b$}}  \put(-95,50){{\color{myblue}$\Pi_c$}}  \put(-5,0){$\pi_1$} \put(-90,75){$\pi_2$}  \put(-5,35){\color{mygr} O6}  \end{picture} & \includegraphics[width=2.8cm]{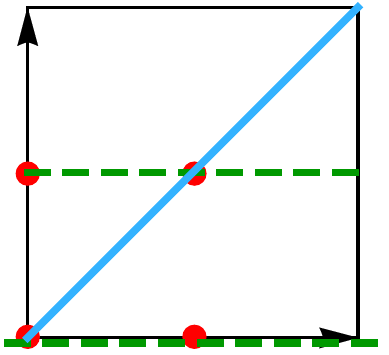}  \begin{picture}(0,0) \put(0,90){$T^4/\Z_2$} \put(-80,-10){{\color{red} 1}}  \put(-45,-10){{\color{red} 2}} \put(-90,35){{\color{red} 3}}  \put(-36,35){{\color{red} 4}} \put(-5,0){$\pi_3$} \put(-90,75){$\pi_6$}    \end{picture} & \includegraphics[width=2.8cm]{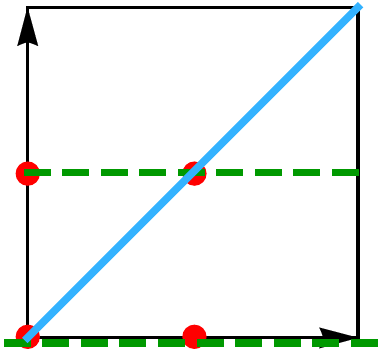}  \begin{picture}(0,0) \put(-80,-10){{\color{red} 1}}  \put(-45,-10){{\color{red} 2}} \put(-90,35){{\color{red} 3}}  \put(-36,35){{\color{red} 4}}  \put(-5,0){$\pi_5$} \put(-90,75){$\pi_4$}  \end{picture}
\end{tabular}
\caption{Geometric representation of the orientifold $(T^2\times K3)/\Omega_p (-)^{F_L} {\cal R}$ in terms of factorised two-tori. The red points $1,2,3$ and $4$ correspond to the $\Z_2$ fixed points, while the green dashed lines indicate the fixed planes under the anti-holomorphic involution ${\cal R}$. The torus wrapping numbers for the three-cycles $\Pi_a$, $\Pi_b$ and $\Pi_c$ are given in equation~(\ref{Eq:D6ModelT2K3}) and the factorisation of $T^4/\Z_2$ has been chosen to easily depict $\Pi_c^{\rm bulk}$. \label{Fig:T2K3Model}}
\end{center}
\end{figure}

Next, we consider three D6-branes stacks $a,b$ and $c$ supported by three different fractional three-cycles $\Pi_a$, $\Pi_b$ and $\Pi_c$ respectively, with torus wrapping numbers given by
\begin{equation}\label{Eq:D6ModelT2K3}
\begin{array}{ll}
\Pi_a:& (1,0)_{(\pi_1,\pi_2)} \times (1,0)_{(\pi_3,\pi_4)}  \times (1,0)_{(\pi_5,\pi_6)}, \\
\Pi_b:& (1,0)_{(\pi_1,\pi_2)} \times (0,1)_{(\pi_3,\pi_4)} \times (0,-1)_{(\pi_5,\pi_6)}, \\
\Pi_c:& (1,0)_{(\pi_1,\pi_2)} \times (1,1)_{(\pi_3,\pi_6)} \times (1,1)_{(\pi_5,\pi_4)}. 
\end{array}
\end{equation}
Note the bulk part of the fractional three-cycle $\Pi_c$ is non-factorisable and should be written as the linear combination $\Pi_c^{\rm bulk} = \pi_{\beta^0} - \pi_{\beta^1} + \pi_1 \otimes \pi_3 \otimes \pi_4  - \pi_1 \otimes \pi_5 \otimes \pi_6$. Such non-factorisable three-cycles can be related \cite{Font:2006na} to coisotropic D8-branes through T-dualities, and share with the latter the property that the lagrangian condition $J|_{\Pi_c} \neq 0$ is violated for anisotropic untwisted K\"ahler moduli, i.e.~$T^2\neq T^3$. This potential violation is the source for the bilinear coupling in the open-closed D6-brane superpotential in this toy model. Since the non-factorisable three-cycles $\pi_1 \otimes \pi_3 \otimes \pi_4$ and  $\pi_1 \otimes \pi_5 \otimes \pi_6$ are ${\cal R}$-odd under the orientifold projection, they do not yield non-vanishing RR tadpoles.
Each fractional three-cycle is frozen at four separate fixed points $(ij)$ on $T^4/\Z_2$, such that the D6-brane position moduli along $T^4/\Z_2$ are projected out. These fractional three-cycles thus only retain the position-moduli $\Phi^1_{\alpha=a,b,c}$ along the first two-torus, as indicated in figure~\ref{Fig:T2K3Model} by virtue of arrows. Furthermore, the exceptional part for each fractional three-cycle forms a linear combination of four ${\cal R}$-odd exceptional basis cycles $\varepsilon_{ij}$, such that the sum of a fractional three-cycle with its orientifold image only wraps bulk three-cycle generators. This consideration implies that the twisted RR tadpoles automatically vanish and drastically simplifies the bulk RR tadpole cancelation conditions (\ref{RRtadpole}),
\begin{equation}
N_a + N_c = 16, \qquad N_b +N_c = 16,
\end{equation} 
which are satisfied for the gauge group choice $SO(32-2N)_a\times USp(32-2N)_b\times U(N)_c$. The $SO(2N_a)$ enhancement of the $a$-stack gauge group follows when the $a$-stack lies on top of the $\Omega {\cal R}$-plane along $T_{(1)}^2$, while the $USp(2N_b)$-enhancement of the $b$-stack gauge group occurs for a $b$-stack on top of the $\Omega {\cal R}\Z_2$-plane along $T_{(1)}^2$. In case the cycles are pulled away from the O6-planes along $T_{(1)}^2$, the gauge group supported by the three fractional three-cycles corresponds to $U(16-N)_a\times U(16-N)_b\times U(N)_c$. Given that all three fractional cycles are parallel to the $\Omega {\cal R}\Z_2$-plane along the first two-torus $T_{(1)}^2$, the K-theory constraints are trivially satisfied.  

For this consistent D6-brane configuration, we can now infer the structure of the K\"ahler potential and superpotential. We therefore introduce local coordinates $z^i = x^i + i\, y^i$ on each two-torus $T_{(i)}^2$ with periodicity $x^i \sim x^i +1$ and $y^i \sim y^i +1$ along the basis one-cycles. In first instance, we compute the quantities $({\cal Q}_\alpha^K)_{ij}$ defined in (\ref{Qs}) by identifying the harmonic one-form $\l_s^{-1}\zeta_1 = dx^1 \in {\cal H}^{1}(\Pi_\alpha^0, \Z)$  compatible with the D-brane normal deformation $X = \oh l_s\partial_{y^1}$ parallel to $\pi_2$ for all fractional three-cycles. An explicit computation for the three D6-brane stacks then shows:
\begin{equation}
\begin{array}{l@{\hspace{0.4in}}l}
({\cal Q}_a^0)_{11} = 0, & ({\cal Q}_a^1)_{11} = -1, \\
({\cal Q}_b^0)_{11} = 1, & ({\cal Q}_b^1)_{11} = 0, \\
({\cal Q}_c^0)_{11} = 1, & ({\cal Q}_c^1)_{11} = -1. \\
\end{array}
\end{equation}
The components $({\cal Q}_\alpha^{K=2,3})_{11}$ vanish as the interior products of $\beta^{K=2,3}$ with respect to $X^1$ vanish, implying that only the complex structure moduli $N^0$ and $N^1$ will be redefined by the open string moduli $\Phi^1_\alpha$. The rigidity of the fractional three-cycles along $T^4/\Z_2$ also implies that components $({\cal Q}_\alpha^{K})_{ij}$ with $i,j \in \{2,3\}$ vanish. Secondly, we have to compute the quantities $(\eta_{\a\, a}^0)^i{}_j$ as defined in (\ref{etas}). To this end, we introduce the harmonic two-form $l_s^{-2}\rho^1_\alpha \in {\cal H}^2 (\Pi_\alpha^0,\Z)$  for each cycle $\alpha \in \{a,b,c\}$, such that it forms the Poincar\'e dual to $\zeta^1$ on $\Pi_\alpha^0$. A straightforward computation then reveals: 
\begin{equation}
(\eta_{a\, 1}^0)^1{}_1=- 1, \qquad (\eta_{b\, 1}^0)^1{}_1=-1, \qquad (\eta_{c\, 1}^0)^1{}_1=-1,
\end{equation}
while all other components vanish. Hence, the full K\"ahler potential for the bulk moduli is given by: 
\begin{equation}
\begin{array}{rcl}
K_Q + K_K &=& - \log \left( N^{0} - \ov N^{0}  - \frac{1}{4}  \frac{ ( \Phi^1_b - \ov  \Phi^1_b )^2 + (\Phi^1_c - \ov \Phi^1_c )^2 }{T^1 - \ov T^1}    \right) \\
&& - \log \left( N^{1} - \ov N^{1}  + \frac{1}{4}  \frac{ (\Phi^1_a - \ov \Phi^1_a)^2 + ( \Phi^1_c - \ov \Phi^1_c)^2 }{ T^1 - \ov T^1}   \right)  \\
&& -  \sum_{k=2}^3\log \left( N^k - \ov N^k \right) - \sum_{i=1}^3 \log \left(i (T^i - \ov T^i) \right) ,
\end{array}
\end{equation}
where we used the redefinition (\ref{redefTorus}) for the complex structure moduli $N'^{0}$ and $N'^{1}$.
And lastly, in order to determine the open-closed superpotential ${W}_{\rm D6}$ in (\ref{supoD6full}) for this D6-brane configuration we have to calculate the geometric quantities  $n_{ai}^\alpha$ defined through:
\begin{equation}
n_{a1}^\alpha = l_s^{-3} \int_{\Pi_\alpha} \omega_a \wedge \zeta_1.
\end{equation}
Keeping in mind the special Lagrangian condition for the fractional three-cycles enables us to pull-back the two-form $\omega_\alpha$ with respect to $\Pi_\alpha$ and to obtain the geometric quantities $n_{ai}^\alpha$:
\begin{equation}
n^c_{11}=0, \quad n^c_{21}=1 = - n^c_{31}, \qquad  n_{ai}^\alpha = 0 \quad \alpha \in \{a,b\} \, (\forall\, a, i).
 \end{equation}
The full open-closed superpotential ${W}_{\rm D6}$, including a D-brane worldvolume flux $F$ supported by the de Rahm dual two-cycles to the two-forms $\rho^1_\alpha$, thus reads:
\begin{equation}\label{Eq:OCSToyModel}
l_s {W}_{\rm D6} = -   \Phi_c^1  (T^2 - T^3).
\end{equation}
Given the K\"ahler potential and the open-closed superpotential one can determine the scalar potential explicitly along the lines of section~\ref{s:spot}. The RR part of the scalar potential $V_{RR+CS}$ is given by (\ref{VRRCS}) with the redefined flux quanta given by:  
\begin{equation}
\begin{array}{rcl@{\hspace{0.4in}}rcl}
\tilde e_0 &=& e_0 &   \\
\tilde e_1 &=& e_1 , &\tilde m^1 &=& m^1,\\
\tilde e_2 & =& e_2 -  \theta^1_c, & \tilde m^2 &=& m^2 + \varphi_c^1 \\
\tilde e_3 & =& e_3 +   \theta^1_c, & \tilde m^3 &=& m^3 - \varphi_c^1. \\
\end{array}
\end{equation}
The contributions to the redefined flux quanta $\tilde m^a$ can be easily computed using equation (\ref{pullbackid}) and the quantities $n_{ai}^\alpha$ and $(\eta_{\a\, a}^0)^i{}_j$ calculated above. The DBI-part follows from expression (\ref{VDBI}) and the open-closed superpotential (\ref{Eq:OCSToyModel}):
\begin{equation}
V_{\rm DBI}  =  \frac{e^K}{l_s^2 \kappa_4^2} G^{11}_{{\rm D6}_c} \left(  T^2 - T^3\right)\left( \ov{T}^2 - \ov{T}^3\right)  .
\end{equation}
The inverse metrics $G^{11}_{{\rm D6}_\alpha}$ on the open string moduli spaces have to be determined for each D6-brane $\Pi_\alpha$ separately and depend implicitly on the K\"ahler modulus $T^1$ and the complex structure moduli $N^0$ and $N^1$.


\begin{thebibliography}{10}


\bibitem{Ibanez:2012zz} 
  L.~E.~Iba\~nez and A.~M.~Uranga,
  {\em ``String theory and particle physics: An introduction to string phenomenology,''}
  Cambridge University Press (2012).
 
   \bibitem{Grana:2005jc} 
  M.~Gra\~na,
  {\em ``Flux compactifications in string theory: A Comprehensive review,''}
  Phys.\ Rept.\  {\bf 423}, 91 (2006)
  [hep-th/0509003].

\bibitem{Douglas:2006es} 
  M.~R.~Douglas and S.~Kachru,
  {\em ``Flux compactification,''}
  Rev.\ Mod.\ Phys.\  {\bf 79}, 733 (2007)
  [hep-th/0610102].
 
\bibitem{Denef:2007pq} 
  F.~Denef, M.~R.~Douglas and S.~Kachru,
  {\em ``Physics of String Flux Compactifications,''}
  Ann.\ Rev.\ Nucl.\ Part.\ Sci.\  {\bf 57}, 119 (2007)
  [hep-th/0701050].
 
\bibitem{Blumenhagen:2005mu}
  R.~Blumenhagen, M.~Cveti\v c, P.~Langacker, G.~Shiu,
  {\em ``Toward realistic intersecting D-brane models,''}
  Ann.\ Rev.\ Nucl.\ Part.\ Sci.\  {\bf 55 } (2005)  71-139.
  [hep-th/0502005].

\bibitem{Blumenhagen:2006ci}
  R.~Blumenhagen, B.~K\"ors, D.~L\"ust, S.~Stieberger,
  {\em ``Four-dimensional String Compactifications with D-Branes, Orientifolds and Fluxes,''}
  Phys.\ Rept.\  {\bf 445 } (2007)  1-193.
  [hep-th/0610327].

\bibitem{Marchesano:2007de}
  F.~Marchesano,
  {\em ``Progress in D-brane model building,''}
  Fortsch.\ Phys.\  {\bf 55 } (2007)  491-518.
  [hep-th/0702094].

  \bibitem{Nilles:2008gq} 
  H.~P.~Nilles, S.~Ramos-Sanchez, M.~Ratz and P.~K.~S.~Vaudrevange,
  {\em ``From strings to the MSSM,''}
  Eur.\ Phys.\ J.\ C {\bf 59}, 249 (2009)
  [arXiv:0806.3905 [hep-th]].

  
\bibitem{Maharana:2012tu} 
  A.~Maharana and E.~Palti,
  {\em ``Models of Particle Physics from Type IIB String Theory and F-theory: A Review,''}
  Int.\ J.\ Mod.\ Phys.\ A {\bf 28}, 1330005 (2013)
  [arXiv:1212.0555 [hep-th]].

\bibitem{Schellekens:2013bpa}
  A.~N.~Schellekens,
  {\em ``Life at the Interface of Particle Physics and String Theory,''}
  Rev.\ Mod.\ Phys.\  {\bf 85}, no. 4, 1491 (2013)
  [arXiv:1306.5083 [hep-ph]].

\bibitem{Quevedo:2014xia} 
  F.~Quevedo,
  {\em ``Local String Models and Moduli Stabilisation,''}
  arXiv:1404.5151 [hep-th].


\bibitem{Jockers:2008pe} 
  H.~Jockers and M.~Soroush,
  {\em ``Effective superpotentials for compact D5-brane Calabi-Yau geometries,''}
  Commun.\ Math.\ Phys.\  {\bf 290}, 249 (2009)
  [arXiv:0808.0761 [hep-th]].

\bibitem{Grimm:2008dq} 
  T.~W.~Grimm, T.~W.~Ha, A.~Klemm and D.~Klevers,
  {\em ``The D5-brane effective action and superpotential in N=1 compactifications,''}
  Nucl.\ Phys.\ B {\bf 816}, 139 (2009)
  [arXiv:0811.2996 [hep-th]].
  
\bibitem{Baumgartl:2008qp} 
  M.~Baumgartl and S.~Wood,
  {\em ``Moduli Webs and Superpotentials for Five-Branes,''}
  JHEP {\bf 0906}, 052 (2009)
  [arXiv:0812.3397 [hep-th]].

 \bibitem{Anderson:2010mh} 
  L.~B.~Anderson, J.~Gray, A.~Lukas and B.~Ovrut,
  {\em``Stabilizing the Complex Structure in Heterotic Calabi-Yau Vacua,''}
  JHEP {\bf 1102}, 088 (2011)
  [arXiv:1010.0255 [hep-th]].
  
\bibitem{Anderson:2011ty} 
  L.~B.~Anderson, J.~Gray, A.~Lukas and B.~Ovrut,
  {\em ``The Atiyah Class and Complex Structure Stabilization in Heterotic Calabi-Yau Compactifications,''}
  JHEP {\bf 1110}, 032 (2011)
  [arXiv:1107.5076 [hep-th]].

\bibitem{Marchesano:2014iea} 
  F.~Marchesano, D.~Regalado and G.~Zoccarato,
  {\em ``On D-brane moduli stabilisation,''}
  JHEP {\bf 1411}, 097 (2014)
  [arXiv:1410.0209 [hep-th]].

\bibitem{Silverstein:2013wua} 
  E.~Silverstein,
  {\em ``Les Houches lectures on inflationary observables and string theory,''}
  arXiv:1311.2312 [hep-th].

\bibitem{Baumann:2014nda} 
  D.~Baumann and L.~McAllister,
  {\em ``Inflation and String Theory,''}
  arXiv:1404.2601 [hep-th].

\bibitem{Westphal:2014ana} 
  A.~Westphal,
  {\em ``String cosmology Ñ Large-field inflation in string theory,''}
  Int.\ J.\ Mod.\ Phys.\ A {\bf 30}, no. 09, 1530024 (2015)
  [arXiv:1409.5350 [hep-th]].

\bibitem{Bielleman:2015ina} 
  S.~Bielleman, L.~E.~Ib\'a\~nez and I.~Valenzuela,
  {\em ``Minkowski 3-forms, Flux String Vacua, Axion Stability and Naturalness,''}
  JHEP {\bf 1512}, 119 (2015)
  [arXiv:1507.06793 [hep-th]].

\bibitem{Grimm:2011dx} 
  T.~W.~Grimm and D.~V.~Lopes,
  {\em ``The N=1 effective actions of D-branes in Type IIA and IIB orientifolds,''}
  Nucl.\ Phys.\ B {\bf 855}, 639 (2012)
  [arXiv:1104.2328 [hep-th]].

\bibitem{Kerstan:2011dy} 
  M.~Kerstan and T.~Weigand,
  {\em ``The Effective action of D6-branes in N=1 type IIA orientifolds,''}
  JHEP {\bf 1106}, 105 (2011)
  [arXiv:1104.2329 [hep-th]].


\bibitem{Font:2006na} 
  A.~Font, L.~E.~Ib\'a\~nez and F.~Marchesano,
  {\em ``Coisotropic D8-branes and model-building,''}
  JHEP {\bf 0609}, 080 (2006)
  [hep-th/0607219].

\bibitem{Grimm:2004ua} 
  T.~W.~Grimm, J.~Louis,
  {\em ``The Effective action of type IIA Calabi-Yau orientifolds,''}
  Nucl.\ Phys.\  {\bf B718 } (2005) 153-202.
  [hep-th/0412277].


\bibitem{Candelas:1990pi} 
  P.~Candelas and X.~de la Ossa,
  {\em ``Moduli Space of {Calabi-Yau} Manifolds,''}
  Nucl.\ Phys.\ B {\bf 355}, 455 (1991).
  
  \bibitem{Camara:2011jg} 
  P.~G.~C\'amara, L.~E.~Ib\'a\~nez and F.~Marchesano,
  {\em ``RR photons,''}
  JHEP {\bf 1109}, 110 (2011)
  [arXiv:1106.0060 [hep-th]].

\bibitem{Marchesano:2014bia} 
  F.~Marchesano, D.~Regalado and G.~Zoccarato,
  {\em ``U(1) mixing and D-brane linear equivalence,''}
  JHEP {\bf 1408}, 157 (2014)
  [arXiv:1406.2729 [hep-th]].
  
\bibitem{McLean}
  R.~C.~McLean,
  {\em ``Deformations of calibrated submanifolds,''}
  Comm.\ Anal.\ Geom.\ {\bf 6} (1998) 705-747.
 
\bibitem{Cremades:2003qj} 
  D.~Cremades, L.~E.~Ib\'a\~nez and F.~Marchesano,
  {\em ``Yukawa couplings in intersecting D-brane models,''}
  JHEP {\bf 0307}, 038 (2003)
  [hep-th/0302105].

\bibitem{Berg:2011ij} 
  M.~Berg, M.~Haack and J.~U.~Kang,
  {\em ``One-Loop K\"ahler Metric of D-Branes at Angles,''}
  JHEP {\bf 1211}, 091 (2012)
  [arXiv:1112.5156 [hep-th]].
 
 
\bibitem{Hitchin:1997ti} 
  N.~J.~Hitchin,
  {\em ``The Moduli space of special Lagrangian submanifolds,''}
  Annali Scuola Sup.\ Norm.\ Pisa Sci.\ Fis.\ Mat.\  {\bf 25}, 503 (1997)
  [dg-ga/9711002].

\bibitem{Kachru:2000an} 
  S.~Kachru, S.~H.~Katz, A.~E.~Lawrence and J.~McGreevy,
  {\em ``Mirror symmetry for open strings,''}
  Phys.\ Rev.\ D {\bf 62}, 126005 (2000)
  [hep-th/0006047].
 

\bibitem{Louis:2002ny}
  J.~Louis and A.~Micu,
  {\em ``Type 2 theories compactified on Calabi-Yau threefolds in the presence of background fluxes,''}
  Nucl.\ Phys.\ B {\bf 635} (2002) 395
  [hep-th/0202168].


\bibitem{Bergshoeff:2001pv} 
  E.~Bergshoeff, R.~Kallosh, T.~Ortin, D.~Roest and A.~Van Proeyen,
  {\em ``New formulations of D = 10 supersymmetry and D8 - O8 domain walls,''}
  Class.\ Quant.\ Grav.\  {\bf 18}, 3359 (2001)
  [hep-th/0103233].

\bibitem{Villadoro:2005cu} 
  G.~Villadoro and F.~Zwirner,
  {\em ``N=1 effective potential from dual type-IIA D6/O6 orientifolds with general fluxes,''}
  JHEP {\bf 0506}, 047 (2005)
  [hep-th/0503169].

\bibitem{Marolf:2000cb} 
  D.~Marolf,
  {\em ``Chern-Simons terms and the three notions of charge,''}
  hep-th/0006117.

\bibitem{Escobar:2015ckf} 
  D.~Escobar, A.~Landete, F.~Marchesano and D.~Regalado,
    {\em ``Large field inflation from D-branes,''}
  Phys.\ Rev.\ D {\bf 93}, no. 8, 081301 (2016)
  [arXiv:1505.07871 [hep-th]];
  {\em ``D6-branes and axion monodromy inflation,''}
  JHEP {\bf 1603}, 113 (2016)
  doi:10.1007/JHEP03(2016)113
  [arXiv:1511.08820 [hep-th]].

\bibitem{Gukov:1999gr} 
  S.~Gukov,
  {\em ``Solitons, superpotentials and calibrations,''}
  Nucl.\ Phys.\ B {\bf 574}, 169 (2000)
  [hep-th/9911011].

\bibitem{Taylor:1999ii} 
  T.~R.~Taylor and C.~Vafa,
  {\em ``R R flux on Calabi-Yau and partial supersymmetry breaking,''}
  Phys.\ Lett.\ B {\bf 474}, 130 (2000)
  [hep-th/9912152].

\bibitem{Thomas:2001ve} 
  R.~P.~Thomas,
  {\em ``Moment maps, monodromy and mirror manifolds,''}
  math/0104196 [math-dg].
  
\bibitem{Martucci:2006ij} 
  L.~Martucci,
  {\em ``D-branes on general N=1 backgrounds: Superpotentials and D-terms,''}
  JHEP {\bf 0606}, 033 (2006)
  [hep-th/0602129].
  
\bibitem{Marchesano:2014mla} 
  F.~Marchesano, G.~Shiu and A.~M.~Uranga,
  {\em ``F-term Axion Monodromy Inflation,''}
  JHEP {\bf 1409}, 184 (2014)
  [arXiv:1404.3040 [hep-th]].

\bibitem{Dvali:2005ws}
  G.~Dvali, R.~Jackiw and S.~-Y.~Pi,
  {\em ``Topological mass generation in four dimensions,''}
  Phys.\ Rev.\ Lett.\  {\bf 96} (2006) 081602
  [hep-th/0511175].
  
\bibitem{Dvali:2005an}
  G.~Dvali,
  {\em ``Three-form gauging of axion symmetries and gravity,''}
  hep-th/0507215.

 \bibitem{Kaloper:2008fb} 
  N.~Kaloper and L.~Sorbo,
  {\em ``A Natural Framework for Chaotic Inflation,''}
  Phys.\ Rev.\ Lett.\  {\bf 102}, 121301 (2009)
  [arXiv:0811.1989 [hep-th]].
  
  \bibitem{Kaloper:2011jz} 
  N.~Kaloper, A.~Lawrence and L.~Sorbo,
  {\em ``An Ignoble Approach to Large Field Inflation,''}
  JCAP {\bf 1103}, 023 (2011)
  [arXiv:1101.0026 [hep-th]].

\bibitem{Lerche:2001cw} 
  W.~Lerche and P.~Mayr,
  {\em ``On N=1 mirror symmetry for open type 2 strings,''}  
  hep-th/0111113.

\bibitem{Lerche:2002ck} 
  W.~Lerche, P.~Mayr and N.~Warner,
  {\em ``Holomorphic N=1 special geometry of open - closed type II strings,''}
  hep-th/0207259.


\bibitem{Lerche:2002yw} 
  W.~Lerche, P.~Mayr and N.~Warner,
  {\em ``N=1 special geometry, mixed Hodge variations and toric geometry,''}
  hep-th/0208039.



\bibitem{Lerche:2003hs} 
  W.~Lerche,
  {\em ``Special geometry and mirror symmetry for open string backgrounds with N = 1 supersymmetry,''}
  hep-th/0312326.


\bibitem{Koerber:2007xk} 
  P.~Koerber and L.~Martucci,
  {\em ``From ten to four and back again: How to generalize the geometry,''}
  JHEP {\bf 0708}, 059 (2007)
  [arXiv:0707.1038 [hep-th]].

\bibitem{Hitchin:1999fh} 
  N.~J.~Hitchin,
  {\em ``Lectures on special Lagrangian submanifolds,''}
  math/9907034.


\bibitem{Antoniadis:1996vw} 
  I.~Antoniadis, C.~Bachas, C.~Fabre, H.~Partouche and T.~R.~Taylor,
  {\em ``Aspects of type I - type II - heterotic triality in four-dimensions,''}
  Nucl.\ Phys.\ B {\bf 489}, 160 (1997)
  doi:10.1016/S0550-3213(96)00514-7
  [hep-th/9608012].

\bibitem{DeWolfe:2002nn} 
  O.~DeWolfe and S.~B.~Giddings,
  {\em ``Scales and hierarchies in warped compactifications and brane worlds,''}
  Phys.\ Rev.\ D {\bf 67}, 066008 (2003)
  [hep-th/0208123].

\bibitem{Angelantonj:2003zx} 
  C.~Angelantonj, R.~D'Auria, S.~Ferrara and M.~Trigiante,
  {\em ``K3 x T**2 / Z(2) orientifolds with fluxes, open string moduli and critical points,''}
  Phys.\ Lett.\ B {\bf 583}, 331 (2004)
  [hep-th/0312019].

\bibitem{Grana:2003ek} 
  M.~Gra\~na, T.~W.~Grimm, H.~Jockers and J.~Louis,
  {\em ``Soft supersymmetry breaking in Calabi-Yau orientifolds with D-branes and fluxes,''}
  Nucl.\ Phys.\ B {\bf 690}, 21 (2004)
  [hep-th/0312232].

\bibitem{Berg:2004ek} 
  M.~Berg, M.~Haack and B.~K\"ors,
  {\em ``Loop corrections to volume moduli and inflation in string theory,''}
  Phys.\ Rev.\ D {\bf 71}, 026005 (2005)
  [hep-th/0404087].

\bibitem{Jockers:2004yj}
  H.~Jockers and J.~Louis,
  {\em ``The Effective action of D7-branes in N = 1 Calabi-Yau orientifolds,''}
  Nucl.\ Phys.\ B {\bf 705} (2005) 167
  [hep-th/0409098].

 \bibitem{Baumann:2006th} 
  D.~Baumann, A.~Dymarsky, I.~R.~Klebanov, J.~M.~Maldacena, L.~P.~McAllister and A.~Murugan,
  {\em ``On D3-brane Potentials in Compactifications with Fluxes and Wrapped D-branes,''}
  JHEP {\bf 0611}, 031 (2006)
  [hep-th/0607050].
 
 \bibitem{Marchesano:2008rg} 
  F.~Marchesano, P.~McGuirk and G.~Shiu,
  {\em ``Open String Wavefunctions in Warped Compactifications,''}
  JHEP {\bf 0904}, 095 (2009)
  doi:10.1088/1126-6708/2009/04/095
  [arXiv:0812.2247 [hep-th]].
 
\bibitem{Camara:2009uv} 
  P.~G.~C\'amara, C.~Condeescu and E.~Dudas,
  {\em ``Holomorphic variables in magnetized brane models with continuous Wilson lines,''}
  JHEP {\bf 1004}, 029 (2010)
  [arXiv:0912.3369 [hep-th]].

\bibitem{BerasaluceGonzalez:2012vb} 
  M.~Berasaluce-Gonzalez, P.~G.~Camara, F.~Marchesano, D.~Regalado and A.~M.~Uranga,
  {\em ``Non-Abelian discrete gauge symmetries in 4d string models,''}
  JHEP {\bf 1209}, 059 (2012)
  [arXiv:1206.2383 [hep-th]].

\bibitem{Martucci:2014ska} 
  L.~Martucci,
  {\em ``Warping the K\"ahler potential of F-theory/IIB flux compactifications,''}
  JHEP {\bf 1503}, 067 (2015)
  [arXiv:1411.2623 [hep-th]].

\bibitem{Berg:2005ja} 
  M.~Berg, M.~Haack and B.~K\"ors,
  {\em ``String loop corrections to K\"ahler potentials in orientifolds,''}
  JHEP {\bf 0511}, 030 (2005)
  [hep-th/0508043].

\bibitem{Berg:2005yu} 
  M.~Berg, M.~Haack and B.~K\"ors,
  {\em ``On volume stabilization by quantum corrections,''}
  Phys.\ Rev.\ Lett.\  {\bf 96}, 021601 (2006)
  [hep-th/0508171].

\bibitem{Haack:2008yb} 
  M.~Haack, R.~Kallosh, A.~Krause, A.~D.~Linde, D.~L\"ust and M.~Zagermann,
  {\em ``Update of D3/D7-Brane Inflation on K3 x T**2/Z(2),''}
  Nucl.\ Phys.\ B {\bf 806}, 103 (2009)
  [arXiv:0804.3961 [hep-th]].

\bibitem{Berg:2014ama} 
  M.~Berg, M.~Haack, J.~U.~Kang and S.~Sj\"ors,
  {\em ``Towards the one-loop K\"ahler metric of Calabi-Yau orientifolds,''}
  JHEP {\bf 1412}, 077 (2014)
  [arXiv:1407.0027 [hep-th], arXiv:1407.0027].

\bibitem{Grimm:2004uq}
  T.~W.~Grimm and J.~Louis,
  {\em ``The Effective action of N = 1 Calabi-Yau orientifolds,''}
  Nucl.\ Phys.\ B {\bf 699} (2004) 387
  [hep-th/0403067].
  
\bibitem{Grimm:2010ks}
  T.~W.~Grimm,
  ``The N=1 effective action of F-theory compactifications,''
  Nucl.\ Phys.\ B {\bf 845} (2011) 48
  [arXiv:1008.4133 [hep-th]].

\bibitem{Grimm:2015ona}
  T.~W.~Grimm, T.~G.~Pugh and D.~Regalado,
  ``Non-Abelian discrete gauge symmetries in F-theory,''
  JHEP {\bf 1602} (2016) 066
  [arXiv:1504.06272 [hep-th]].

\bibitem{Greiner:2015mdm}
  S.~Greiner and T.~W.~Grimm,
  ``On Mirror Symmetry for Calabi-Yau Fourfolds with Three-Form Cohomology,''
  arXiv:1512.04859 [hep-th].


\bibitem{Giddings:2001yu} 
  S.~B.~Giddings, S.~Kachru and J.~Polchinski,
  {\em ``Hierarchies from fluxes in string compactifications,''}
  Phys.\ Rev.\ D {\bf 66}, 106006 (2002)
  doi:10.1103/PhysRevD.66.106006
  [hep-th/0105097].



\bibitem{Beasley:2002db} 
  C.~Beasley and E.~Witten,
  {\em ``A Note on fluxes and superpotentials in M theory compactifications on manifolds of G(2) holonomy,''}
  JHEP {\bf 0207}, 046 (2002)
  [hep-th/0203061].


\bibitem{Martucci:2009sf} 
  L.~Martucci,
  {\em ``On moduli and effective theory of N=1 warped flux compactifications,''}
  JHEP {\bf 0905}, 027 (2009)
  [arXiv:0902.4031 [hep-th]].


\bibitem{backreaction}
  A.~Landete, F.~Marchesano and C.~Wieck,
  {\em ``Challenges for D-brane large-field inflation with stabilizer fields,''}
  arXiv:1607.01680 [hep-th].

 \bibitem{DeWolfe:2005uu} 
  O.~DeWolfe, A.~Giryavets, S.~Kachru and W.~Taylor,
  {\em ``Type IIA moduli stabilization,''}
  JHEP {\bf 0507}, 066 (2005)
  [hep-th/0505160].
  
  \bibitem{Camara:2005dc} 
  P.~G.~C\'amara, A.~Font and L.~E.~Ib\'a\~nez,
  {\em ``Fluxes, moduli fixing and MSSM-like vacua in a simple IIA orientifold,''}
  JHEP {\bf 0509}, 013 (2005)
  doi:10.1088/1126-6708/2005/09/013
  [hep-th/0506066].
  
  \bibitem{Palti:2008mg} 
  E.~Palti, G.~Tasinato and J.~Ward,
  {\em ``WEAKLY-coupled IIA Flux Compactifications,''}
  JHEP {\bf 0806}, 084 (2008)
  [arXiv:0804.1248 [hep-th]].
  
\bibitem{Gomis:2005wc} 
  J.~Gomis, F.~Marchesano and D.~Mateos,
  {\em ``An Open string landscape,''}
  JHEP {\bf 0511}, 021 (2005)
  doi:10.1088/1126-6708/2005/11/021
  [hep-th/0506179].
  
\bibitem{Marchesano:2006ns} 
  F.~Marchesano,
  {\em ``D6-branes and torsion,''}
  JHEP {\bf 0605}, 019 (2006)
  [hep-th/0603210].
  
\bibitem{Koerber:2006hh} 
  P.~Koerber and L.~Martucci,
  {\em ``Deformations of calibrated D-branes in flux generalized complex manifolds,''}
  JHEP {\bf 0612}, 062 (2006)
  [hep-th/0610044].

\bibitem{Garcia-Valdecasas:2016voz} 
  E.~Garc\'ia-Valdecasas and A.~Uranga,
  {\em ``On the 3-form formulation of axion potentials from D-brane instantons,''}
  arXiv:1605.08092 [hep-th].

\bibitem{Marchesano:2013ega} 
  F.~Marchesano, D.~Regalado and L.~Vazquez-Mercado,
  {\em ``Discrete flavor symmetries in D-brane models,''}
  JHEP {\bf 1309}, 028 (2013)
  [arXiv:1306.1284 [hep-th]].
  



\bibitem{Ferrara:1994kg} 
  S.~Ferrara, C.~Kounnas and F.~Zwirner,
  {\em ``Mass formulae and natural hierarchy in string effective supergravities,''}
  Nucl.\ Phys.\ B {\bf 429}, 589 (1994)
  Erratum: [Nucl.\ Phys.\ B {\bf 433}, 255 (1995)]
  [hep-th/9405188].

\bibitem{Ciupke:2015ora} 
  D.~Ciupke and L.~Zarate,
  {\em ``Classification of Shift-Symmetric No-Scale Supergravities,''}
  JHEP {\bf 1511}, 179 (2015)
  [arXiv:1509.00855 [hep-th]].

\bibitem{Gukov:2002iq}
  S.~Gukov and M.~Haack,
  {\em ``IIA string theory on Calabi-Yau fourfolds with background fluxes,''}
  Nucl.\ Phys.\ B {\bf 639} (2002) 95
  [hep-th/0203267].



\bibitem{Blumenhagen:2002wn}
  R.~Blumenhagen, V.~Braun, B.~K\"ors and D.~L\"ust,
  {\em ``Orientifolds of K3 and Calabi-Yau manifolds with intersecting D-branes,''}
  JHEP {\bf 0207} (2002) 026
  [hep-th/0206038].


  
  

  
  
  
\end{thebibliography}
\end{document}